\documentclass[aps,preprint,amsmath,amssymb,nofootinbib]{revtex4-2}

\usepackage{xcolor}
\usepackage{graphicx}

\begin{document}


\title{Journey to the center of stars: the realm of low Prandtl number fluid dynamics}



\author{P. Garaud$^1$}
\affiliation{
$^{1}$Department of Applied Mathematics, Baskin School of Engineering, \\
University of California, Santa Cruz, CA 95064\\
}



\begin{abstract}
The dynamics of fluids deep in stellar interiors is a subject that bears many similarities with geophysical fluid dynamics, 
with one crucial difference: the Prandtl number. The ratio of the kinematic viscosity to the thermal diffusivity is usually of order unity or more on Earth, but 
is always much smaller than one in stars. As a result, viscosity remains negligible on scales that are thermally diffusive, which opens 
the door to a whole new region of parameter space, namely the turbulent low P\'eclet number regime (where the P\'eclet number is the product of the Prandtl number and the Reynolds number).
In this review, I focus on three instabilities that are well known in geophysical fluid dynamics, and have
an important role to play in stellar evolution, namely convection, stratified shear instabilities, and double-diffusive convection. I present what is known of their behavior at low Prandtl number, highlighting the differences with their moderate and high Prandtl number counterparts. 
\end{abstract}



\maketitle

\section{Introduction}
\label{sec:intro}

\subsection{Stellar structure and evolution (abridged)}

Stars have long fascinated human beings, but our modern understanding of stellar structure and evolution is only about 100 years old. By and large, the vast majority of stars are almost spherical balls of gas in quasi-hydrostatic and thermal equilibrium, the combination of which dictates their structure. Nuclear reactions in the high-density stellar core heat the plasma to extremely high temperatures, therefore maintaining the enormous hydrostatic pressure needed to prevent the star from collapsing gravitationally. The heat is transported from the core to the stellar surface either by radiative diffusion or convection (whichever is most efficient), and from there is radiated into space for us to observe. A star's spectrum can provide a wealth of information about its structure. The spectrum itself is usually quite close to that of a black body, therefore revealing the star's surface temperature. The luminosity depends on the star's emitting surface area, which can be used to measure its radius. Often, individual emission or absorption lines are also observed on top of the black-body spectrum. Their wavelength can reveal the composition of the surface layers of the star, while their width can provide information on the surface gravity and (in some cases) rotation or magnetic fields. 

Of course, stars are not exactly in a steady state, since the nuclear reactions required to heat up the core gradually change their internal composition. However, these reactions are usually much slower than the thermal  evolution timescale, which is why a quasi-static model is appropriate. The most common nuclear fusion reaction transforms hydrogen into helium (but others can also take place depending on the temperature and density within the core). This reaction is extremely stable and slow, and sets the star's aging rate. The star reaches its end-of-life phase after exhausting all the hydrogen in the core. As such, its lifespan is sensitively dependent on the inward flux of hydrogen fuel into the core, from advection by large-scale flows or small-scale turbulent mixing for instance. Note that while these flows can rarely be directly measured, indirect evidence of their presence can sometimes be found in the star's spectrum. Indeed, the same fluid dynamical processes that fuel the core also transport various nuclear fusion or fission products and by-products  (such as lithium, beryllium, boron, carbon, nitrogen, oxygen and their respective isotopes) from the core back to the surface, that can be detected spectroscopically.

Through this highly abridged description of stellar structure and evolution (see, e.g. \cite{CoxGiuli1968,KippenhahnWeigert2012} for more detail), we see that modeling stars relies on accurately accounting for both microphysical and macrophysical transport processes. Statistical mechanics, nuclear physics and quantum mechanics are required to model the plasma's equation of state, compute the nuclear reaction rates, and quantify radiative diffusion coefficients for heat transport. Fluid mechanics, on the other hand, is required to model heat and chemical transport by convection as well as any other possible source of turbulence in convectively stable regions. Thanks to historical events of the 1930s-1970s, scientific understanding of microphysical processes involved in nuclear reactions and radiative heat transport are very well understood. The fluid dynamics of stellar evolution, however, remains relatively poorly constrained in comparison. 

\subsection{Stellar fluid dynamics}

Today, the general consensus is that many of the remaining discrepancies between stellar models and observations can be attributed to inadequate or missing prescriptions for turbulent transport. 
Turbulence in stars arises from two classes of instabilities: thermal convection on the one hand, and all other instabilities of stably stratified fluids on the other hand (see, e.g. the graduate textbooks by \citet{Stix2004} for the solar interior, and by \citet{KatoFukue2020} for stars more generally).
Convection is by far the most important fluid dynamical process in stars. As discussed above, it transports heat outward, and in many cases forces the stratification to become almost adiabatic, thus setting the density structure of the star 
(and therefore its size, given a certain mass). Almost all stars contain a region that is convectively unstable; assuming a chemical composition that is close to solar, lower mass stars ($M_\star \lesssim 0.4 M_\odot$) are fully convective, solar-type stars ($0.4 M_\odot < M_\star < 1.3 M_\odot$) have an outer convection zone surrounding a stably stratified core, intermediate mass stars ($1.3 M_\odot \lesssim M_\star \lesssim 10 M_\odot$) have a convective core surrounded by a stably stratified envelope, and higher mass stars can even have multiple convective regions. As such, convection was, for the longest time, the only dynamical process accounted for in all stellar evolution models. 

Outside of convective regions, by contrast, many different kinds of instabilities can occur. Double-diffusive instabilities, for example, are commonly invoked; these can be of the fingering kind, in regions that are stably stratified in temperature, but unstably stratified in composition, or of the oscillatory or layered kind, in regions that are unstably stratified in temperature but stably stratified in composition (see the reviews \cite{Garaud2018} and \cite{Garaud2020}). Centrifugal instabilities and shear instabilities are also likely prevalent, since stars are known to be the seat of substantial differential rotation (see the review by Zahn \cite{Zahn1974} for instance). By contrast, other fluid instabilities that are commonly studied in geophysics, such as the baroclinic instability and many others, are much less often discussed -- not because they are not important, but rather, because they have not yet been appreciated by the stellar community (with some exceptions, see \cite{SpruitKnobloch1984} for instance). Instead, the focus has been to better understand the vast range of possible magnetohydrodynamic instabilities (see the review by Mestel \cite{Mestel1999}), such as the various types of Tayler instabilities \citep{Tayler11973,Tayler21973}, joint instabilities of magnetic fields and differential rotation (e.g. \cite{DikpatiGilman1999,Callyal2003}), and magnetic buoyancy instabilities \citep{Acheson1979,SchmittRosner1983}. In what follows I shall therefore focus only on three types of instabilities that are of interest to both the geophysical and astrophysical communities: convection, shear instabilities, and double-diffusive instabilities, and ask the simple question: how do these differ between the different settings? 
 
To answer this simple question, one needs to appreciate the similarities and differences between stellar fluids and geophysical fluids. The main difference is not the compressibility of the plasma, or its composition. 
In fact, deep in stellar interiors, fluid flows are sufficiently slow as to be almost incompressible, and the mean free path of photons is sufficiently short that radiative transfer can be approximated by a diffusion term in the heat equation. As such, ignoring magnetic fields (although see Section \ref{sec:magrot}), the governing equations for stellar fluid dynamics deep below the surface are actually identical to those typically used in atmospheric dynamics, namely 
\begin{eqnarray}
\frac{\partial{\bf u}}{\partial t} + {\bf u}\cdot\nabla {\bf u} & = & - \frac{1}{\rho_m} \nabla \tilde{p} + \alpha_T \tilde{T} g {\bf e}_z + \nu \nabla^2 {\bf u}  \label{eq:momentum}, \\
\nabla \cdot {\bf {u}} &=& 0 \label{eq:continuity},  \\
\frac{\partial \tilde{T}}{\partial t} + {\bf u}\cdot\nabla \tilde{T} + \beta_T w   & = & \kappa_T \nabla^2 \tilde{T} \label{eq:heat}, 
\end{eqnarray}
where ${\bf u} = (u,v,w)$ is the velocity field, $\rho_m$ is the mean density of the fluid, $\tilde{p}$ is the pressure perturbation away from hydrostatic equilibrium, $\tilde{T}$ is the temperature perturbation away from radiative equilibrium, $\alpha_T$ is the coefficient of thermal expansion, $g$ is gravity,  $\nu$ is the kinematic viscosity, $\beta_T = dT_{rad}/dz + g/c_p$ is the potential temperature gradient (where $dT_{rad}/dz$ is the temperature gradient the star would have if it were in radiative equilibrium and $c_p$ is the specific heat at constant pressure), and $\kappa_T$ is the thermal diffusivity. This set of equations was formally derived by \citet{SpiegelVeronis1960}, and generalizes the Boussinesq approximation to weakly compressible fluids.  It is valid as long as the height of the stellar region modeled is much smaller than the local density scaleheight, and the flow velocities are much smaller than the local sound speed. Close to the surface, both the sound speed and the density scaleheight become quite small, and the approximation is not valid. Further down in the interior, however, the density scaleheight is similar to the star's radius and the sound speed increases rapidly with the increasing temperature, so the Spiegel-Veronis-Boussinesq approximation  holds. 

I would therefore argue that, other than the presence/absence of topography and magnetic fields, the only really fundamental difference between stellar and geophysical fluid dynamics is the Prandtl number $Pr = \nu/\kappa_T$, which is typically of order unity or larger in geophysical flows, but is asymptotically small in stellar interiors. This is because radiative diffusion greatly increases thermal transport but only adds a small contribution to momentum transport, so $\kappa_T \gg \nu$ in stars, and $Pr$ ranges between $10^{-9}$ and $10^{-5}$ at most\footnote{Except in cases where the plasma is degenerate, which is a much less frequent situation. Even then, $Pr \sim 10^{-2}$.}. With $Pr = O(1)$ or $Pr \gg 1$, a turbulent flow is always close to adiabatic (i.e. thermally non-diffusive). When $Pr \ll 1$ on the other hand, thermal diffusion can become important even when viscosity is negligible, and this fundamentally changes the nature of fluid instabilities and associated turbulence, as I demonstrate below. As such, there is no reason to expect that any of the well-accepted models for turbulent transport in geophysical flows apply in stellar interiors, and as I shall demonstrate, almost none of them do. This is particularly true for the three types of instabilities highlighted earlier, namely convection, double-diffusive convection, and shear instabilities. 

In what follows, I will first briefly introduce historically-relevant ideas about thermal convection at low Prandtl number in Section \ref{sec:lowPrapprox}, before moving on to instabilities of stably stratified regions. Since the latter often (but not always) have a low P\'eclet number (where the P\'eclet number is the ratio of the thermal diffusive timescale to the thermal advection timescale), it is this property rather than the low Prandtl number that defines their behavior, as shown in Section \ref{sec:LPNapprox}. Section \ref{sec:stratshear} then discusses in turn both vertical and horizontal shear instabilities, and Section \ref{sec:DDC} summarizes our work on double-diffusive instabilities, in both cases at low Prandtl number. Section \ref{sec:magrot} provides a very brief discussion on the need to include rotation and magnetic fields when modeling astrophysical flows, and Section \ref{sec:ccl} concludes with a reflection on the importance of multidisciplinary work, and the fundamental role that the Woods Hole Geophysical Fluid Dynamics (GFD) summer program has played in transforming the field of stellar fluid dynamics. 
 
\section{Thermal convection at low Prandtl number}
\label{sec:lowPrapprox}

As discussed earlier, thermal convection is the only fluid dynamical process that is accounted for in {\it all} modern stellar evolution models. The standard model for convection in stars is called "mixing-length theory", and in its present form originated from the work of Erika B\"ohm-Vitense \citep{Vitense1953}. The model is, at heart, very simple. In the bulk of the convection zone, a rising or sinking parcel of fluid is assumed to travel coherently for one mixing length $l$ before disintegrating and mixing its heat content with the background. The convective potential temperature flux carried by the parcel is $F_{conv} \sim v_{conv} l |\beta_T |$ (noting that $\beta_T < 0$ for convective regions), where the parcel velocity is estimated by balancing its kinetic and potential energy, to be $v_{conv} \sim \sqrt{\alpha_T g |\beta_T|  l^2 }$. Combining the two leads to 
\begin{equation}
F_{conv} \sim \sqrt{\alpha_T g} l^2 |\beta_T |^{3/2} .
\label{eq:Fconv}
\end{equation}
By introducing a Rayleigh number based on the mixing length $l$ as $Ra_l = \alpha_T | \beta_T | g l^4/(\kappa_T \nu)$, we see that 
\begin{equation}
F_{conv} \sim  - (Ra_l Pr)^{1/2} \beta_T \kappa_T = (Ra_l Pr)^{1/2}  \frac{\beta_T}{dT_{rad}/dz} F_{diff} ,
\end{equation} 
where $F_{diff} = - \kappa_T dT_{rad}/dz$ is the diffusive temperature flux. This implies that, within the context of the mixing-length model for stellar convection, the Nusselt number (i.e. the ratio of convective to diffusive fluxes) scales as
\begin{equation}
Nu \sim (Ra_l Pr)^{1/2}.
\label{eq:vitense}
\end{equation}
Mixing-length theory has been used by stellar astrophysicists for over sixty years to model stellar convection, with the mixing length $l$ usually chosen to be a fraction of the local pressure scaleheight. Additional physics are invoked to constrain the pre-factors, deal with complex boundary conditions and implement the convective flux prescription (\ref{eq:Fconv}) in stellar evolution codes, but these do not modify the main result substantially. 

Meanwhile, around the same time as Vitense's work, \citet{Priestley1954} and \citet{Malkus1954} were attacking the problem of heat transport in Rayleigh-B\'enard convection (i.e. thermal convection between two parallel plates held at different temperatures, RBC hereafter), with geophysical applications in mind. Both independently came to the conclusion that the Nusselt number in RBC should scale as 
\begin{equation}
Nu \sim Ra^{1/3}, 
\end{equation}
where here
\begin{equation}
Ra = \frac{\alpha_T | \beta_T | g H^4}{\kappa_T \nu}, 
\label{eq:RaRBC}
\end{equation}
and $H$ is the distance between the plates. 
Priestley's derivation relies on dimensional analysis, while Malkus' treatment assumes that the turbulence would organize itself in such a way as to maximize the heat transport. A similar scaling law was obtained again later by \citet{Howard1966}, who considered the stability of the thermal boundary layers near the plates. At a cursory glance, laboratory experiments (see e.g. \cite{niemelasreenivasan2003}) seem to provide evidence for the 1/3 power law for RBC at least for Rayleigh numbers up to $10^{14}$, albeit for fluids that have an $O(1)$ Prandtl number. Whether the $Nu$ vs $Ra$ scaling deviates from this law above $10^{14}$ or not is currently controversial \citep{Heal2012,Doering2020}. 

As a young astrophysicist in the 1950s, Edward Spiegel was keenly interested in convection, which was the subject of his PhD thesis. Encouraged by George Batchelor to meet Willem Malkus, Spiegel rapidly realized that the geophysical community was leaps and bounds ahead of the astrophysical community in terms of understanding and modeling turbulent flows. He then began to collaborate with many in that field (co-founding the Woods Hole GFD program with Stommel, Malkus, Veronis, Stern, Howard, and Keller), and in the process became one of the very first astrophysicists to apply modern (rigorous, nonlinear) fluid dynamical techniques to the study of astrophysical flows. 

Being aware of both Vitense's and Malkus' work, Spiegel noted that they were incompatible at low Prandtl number, and concluded that Malkus' theory does not apply in that limit \citep{Spiegel1962}. Instead, he proposed  that $Nu \propto (RaPr)^{1/2}$ in that case (while not explicit, this scaling is implied in \cite{Spiegel1963}), which can easily be recovered using the domain scale $H$ for the mixing length $l$ in Vitense's argument, see (\ref{eq:vitense}). A similar result (with additional logarithmic corrections in $Ra$) was obtained by \citet{Kraichnan1962} in the limit of very low Prandtl number and very high Rayleigh number, assuming that the thermal boundary layer near the rigid wall is turbulent instead of being laminar. Today, the $Nu \propto Ra^{1/2}$ law has become known as the Ultimate Regime in RBC and is discussed in both low and moderate $Pr$ limits. Experimentally, the Ultimate Regime has been rather elusive. Laboratory experiments at low Prandtl number are notoriously difficult, and achieving very high Rayleigh numbers is a serious engineering challenge. However, recent works appear to validate the Ultimate Regime scaling, in RBC with rough boundaries (for which the boundary layer becomes turbulent at lower Rayleigh numbers, see \cite{Rocheal2001}) and in internally-heated convection (where the heat source is detached from the boundaries, see \cite{Lepotal2018,Bouillautal2019,Miquelal2020}). Whether this scaling would emerge in standard RBC at low Prandtl number remains to be determined. 

In an attempt to attack the problem more formally \citet{Spiegel1962} proposed the first asymptotic model for convection at very low Prandtl number using an expansion of the Spiegel-Veronis-Boussinesq equations in the limit of $Pr \ll 1$ (see also \cite{Thual1992}). Considering a fluid flow between two parallel horizontal plates held at fixed temperatures $T_m + \Delta T / 2$ and $T_m - \Delta T/2$, separated by a distance $H$, he non-dimensionalized equations  (\ref{eq:momentum})-(\ref{eq:heat}) using the unit length $H$, the unit time $H^2 / \nu$ (which is the viscous timescale across the domain), the unit velocity $\nu/H$ and the unit temperature $(\kappa_T \nu)/(\alpha_T g H^3)$, to get
\begin{eqnarray}
\frac{\partial \hat{\bf u}}{\partial t} + \hat{\bf u}\cdot\nabla \hat{\bf u} & = & - \nabla \hat{p} + Pr^{-1} \hat{T} {\bf e}_z + \nabla^2 \hat {\bf u}  \label{eq:spiegelmomentum}, \\
\nabla \cdot \hat{\bf u} &=& 0 \label{eq:spiegelcontinuity},  \\
\frac{\partial \hat{T}}{\partial t} + \hat {\bf u}\cdot\nabla \hat{T} - Ra \hat{w}  & = & Pr^{-1} \nabla^2 \hat{T} \label{eq:spiegelheat}, 
\end{eqnarray}
where $Ra$ is given by (\ref{eq:RaRBC}), recalling that $\beta_T  < 0$ for RBC. Spiegel then assumed that one may expand each of the dependent variables as a power series in the Prandtl number, namely 
\begin{eqnarray}
\hat T = \hat T_0 + Pr \hat T_1 + ... , \\
\hat {\bf u} = \hat {\bf u} _0 + Pr \hat {\bf u}_1 + ... \quad . \label{eq:uexp}
\end{eqnarray}
Substituting this in (\ref{eq:spiegelmomentum})-(\ref{eq:spiegelheat}), we have, order by order, 
\begin{eqnarray}
\hat T_0 = 0,  \label{eq:lowPr1} \\
\frac{\partial \hat{\bf u}_0}{\partial t} + \hat{\bf u}_0\cdot\nabla \hat{\bf u}_0 & = & - \nabla \hat{p} +  \hat{T}_1 {\bf e}_z + \nabla^2 \hat {\bf u}_0, \label{eq:lowPr2}  \\
- Ra \hat{w}_0  & = &  \nabla^2 \hat{T}_1 \label{eq:lowPr3}, 
\end{eqnarray}
assuming that $\hat {\bf u}_0$ remains $O(1)$ in the $Pr$ expansion. 

These reduced equations, if valid, clearly show that the dynamics of low Prandtl number thermal convection must be fundamentally different from those of standard $Pr = O(1)$ convection. First and foremost, (\ref{eq:lowPr1}) shows that temperature fluctuations must be small, namely $O(Pr)$, which implies that this type of convection cannot affect the background temperature profile to lowest order. This finding is consistent with the reduced expression for the temperature equation (\ref{eq:lowPr3}), in which the convective heat flux (which would normally arise from nonlinearities in the temperature equation) is absent at this order. Taken together, we see that the mechanism usually thought to be responsible for the saturation of RBC at $Pr = O(1)$ (i.e. the modification of the linear background temperature profile by the convective flux into one which has a reduced gradient in the core of the fluid, and an enhanced gradient in thin thermal boundary layers, see, e.g. \cite{MalkusVeronis1958}), cannot operate here. Instead, the nonlinear saturation must proceed through the only remaining nonlinearities, which are in the momentum equation. As such, it is clear that Malkus' theory for heat transport in RBC \cite{Malkus1954}, which relies on arguments of maximization of the convective flux, cannot apply within the context of the low Prandtl number approximation. This is also true for Howard's model \citep{Howard1966}, which relies on the presence of diffusive thermal boundary layers surrounding an almost adiabatic region -- this is not possible here, since the mean stratification must remain almost linear.  

Going back to the asymptotic equations, solving for $\hat T_1$ and substituting it back into the momentum equation (\ref{eq:lowPr2}) leads to 
\begin{equation}
\frac{\partial \hat{\bf u}_0}{\partial t} + \hat{\bf u}_0\cdot\nabla \hat{\bf u}_0 =  - \nabla \hat{p} - Ra \nabla^{-2} \hat w_0  {\bf e}_z + \nabla^2 \hat {\bf u}_0  .
\label{eq:lPrN}
\end{equation}
This shows that the only relevant input parameter characterizing the {\it flow} in the low Prandtl number approximation is the Rayleigh number $Ra$, independently of the Prandtl number. Dominant balance between the nonlinear terms and the buoyancy term (for a more rigorous approach, see \citet{Spiegel1962}) implies that the typical nondimensional flow velocity $\hat w_{rms}$ should be proportional to $Ra$ in this limit. Moreover, since $\hat T = Pr \hat T_1 = - RaPr \nabla^{-2}\hat w$, we then find that the typical temperature fluctuations $\hat T_{rms}$ should be proportional to $Ra^2 Pr$. Finally, these can be used to estimate a Nusselt number as
\begin{equation}
Nu \sim \frac{\hat w_{rms}\hat T_{rms}}{RaPr^{-1}}  \propto (Ra Pr)^2 ,
\label{eq:NuRaSpiegel}
\end{equation}
(see also \cite{Ledouxal1961,Kraichnan1962}). 

\citet{Spiegel1962}, however, immediately realized that there are serious limitations to the applicability of the low Prandtl number equations, when applied to thermal convection. Indeed, this scaling law appears to be at odds with the existence of a formal upper bound to the Nusselt number of the form $Nu < CRa^{1/2}$ (uniformly in $Pr$), see \cite{Howard1963,DoeringConstantin1996}. This implies that equation (\ref{eq:NuRaSpiegel}) must break down at large $Ra$, and as discussed by Spiegel, this is indeed the case. To see why, recall that the low Prandtl number approximation is only valid provided the velocities are  $o(Pr^{-1})$. Since $\hat w_{rms} \propto Ra$, it follows that we need $Ra = o(Pr^{-1})$, while at the same time satisfying $Ra > Ra_c \sim O(10^3)$, the critical threshold for the onset of convection.  This means that unless $Pr$ is really minuscule, the regime of applicability of Spiegel's low Prandtl number approximation for thermal convection is very limited. Once $Ra$ becomes $O(Pr^{-1})$ or larger, the approximation breaks down, and one presumably recovers the mixing length theory scaling $Nu \propto (RaPr)^{1/2}$ \cite{Spiegel1963}. 

As such, it remains unclear whether Spiegel's low Prandtl number approximation \cite{Spiegel1962} is useful to model stellar convection. 
And yet, the legacy of his asymptotic approach to studying low Prandtl number fluids lives on, with once crucial modification -- as discussed below. 


\section{Ligni\`eres' low P\'eclet number approximation}
\label{sec:LPNapprox}

One of the main reasons for the limited applicability of Spiegel's low Prandtl number equations for thermal convection is that turbulent velocities rapidly become very large as the Rayleigh number increases, hence invalidating the assumption that these should be $o(Pr^{-1})$ in (\ref{eq:uexp}). In 1999, Francois Ligni\`eres independently rediscovered these equations, but further noted that they should remain valid for a much wider range of parameter space when applied to model the dynamics of stably stratified fluids, where an increase in stratification tends to lower the turbulent velocities. Crucially, he also argued that the correct expansion parameter ought to be the P\'eclet number instead of the Prandtl number \cite{Lignieres1999}. This is perhaps obvious in hindsight: if the goal of the expansion is to neglect the convective terms in the thermal energy equation, then the latter must be smaller than the desired dominant balance in the equation, namely that between the advection of the background potential temperature, and diffusion. For this to be the case, the ratio between the convective term ${\bf u} \cdot \nabla \tilde T$ and the diffusion term $\kappa_T \nabla^2 \tilde T$ must be small, so we need 
\begin{equation}
Pe_t  = \frac{U_{rms} l}{\kappa_T} \ll 1 ,
\end{equation} 
where $U_{rms}$ is the rms velocity of the flow, and $l$ here is the typical scale of the turbulent eddies. Ligni\`eres' derivation of the {\it low P\'eclet number} approximation therefore begins with normalizing equations (\ref{eq:momentum})-(\ref{eq:heat}) using $U_{rms}$ as the velocity scale, $l$ as the flow scale, and $\beta_T l$ as the temperature scale (assuming $\beta_T > 0$ this time since the fluid is stably stratified), which leads to 
\begin{eqnarray}
\frac{\partial \hat{\bf u}}{\partial t} + \hat{\bf u}\cdot\nabla \hat{\bf u} & = & - \nabla \hat{p} + B \hat{T} {\bf e}_z + Re_t^{-1} \nabla^2 \hat {\bf u}  \label{eq:lignieresmomentum}, \\
\nabla \cdot \hat{\bf u} &=& 0 \label{eq:lignierescontinuity},  \\
\frac{\partial \hat{T}}{\partial t} + \hat {\bf u}\cdot\nabla \hat{T} + \hat{w}  & = & Pe_t^{-1} \nabla^2 \hat{T} \label{eq:lignieresheat}, 
\end{eqnarray}
where $B = \bar N^2 l^2 / U_{rms}^2$ is a buoyancy parameter (the square of an inverse Froude number), and $\bar N = \sqrt{\alpha_T \beta_T g}$ is the usually-defined buoyancy frequency associated with the background stratification. Assuming, in the same spirit as \citet{Spiegel1962}, that 
\begin{eqnarray}
\hat T = \hat T_0 + Pe_t \hat T_1 + ..., \\
\hat {\bf u} = \hat {\bf u} _0 + Pe_t  \hat {\bf u}_1 + ...,
\end{eqnarray}
then, order by order, we have 
 \begin{eqnarray}
\hat T_0 = 0, \label{eq:lowPe1} \\
\frac{\partial \hat{\bf u}_0}{\partial t} + \hat{\bf u}_0\cdot\nabla \hat{\bf u}_0 & = & - \nabla \hat{p} + BPe_t  \hat{T}_1 {\bf e}_z +Re_t^{-1} \nabla^2 \hat {\bf u}_0, \label{eq:lowPe2} \\
\hat{w}_0  & = &  \nabla^2 \hat{T}_1 \label{eq:lowPe3}.
\end{eqnarray}
As in \citet{Spiegel1962}, Ligni\`eres finds that the lowest order temperature fluctuations must be zero (see \ref{eq:lowPe1}), which demonstrates that the turbulence cannot cause any large deviations of $T$ from the background state. Consistent with that, the convective terms again disappear from the thermal energy equation (\ref{eq:lowPe3}). This approximation is only valid provided $\hat {\bf u}_0 \sim O(1)$, but crucially, this condition is now implicitly satisfied from the non-dimensionalization selected. We then have, successively, 
\begin{eqnarray}
\hat {\bf u} & = &  \hat {\bf u}_0,  \\ 
 \hat{T}_1 & = & \nabla^{-2} \hat{w}_0 = \nabla^{-2} \hat w   ,  \label{eq:LPNbalancenondim}  \\ 
\frac{\partial \hat{\bf u}}{\partial t} + \hat{\bf u}\cdot\nabla \hat{\bf u} & = & - \nabla \hat{p} + B Pe_t   \nabla^{-2} \hat w  {\bf e}_z + Re_t^{-1} \nabla^2 \hat {\bf u}.
\end{eqnarray}
This shows that the temperature and the velocity fluctuations are again intimately related to one another. Crucially, this causes a reduction in the dimensionality of parameter space whereby $B$ and $Pe_t$ only ever appear together. As such, stratified flows at low turbulent P\'eclet number only depend on two quantities: the product $B Pe_t$, and the turbulent Reynolds number $Re_t$. Going back to the original dimensional system, we have 
\begin{equation}
\beta_T w  =  \kappa_T \nabla^2 \tilde{T}, 
\label{eq:LPNbalancedim}
\end{equation}
and so the low P\'eclet number approximation leads to
\begin{eqnarray}
\frac{\partial{\bf u}}{\partial t} + {\bf u}\cdot\nabla {\bf u} & = & - \frac{1}{\rho_m} \nabla \tilde{p} + \frac{\bar N^2}{\kappa_T} \nabla^{-2} w  {\bf e}_z + \nu \nabla^2 {\bf u}  \label{eq:LPNmomentum}, \\
\nabla \cdot {\bf {u}} &=& 0 \label{eq:LPNcontinuity},
\end{eqnarray}
showing that the dimensional combination of parameters $ \bar N^2 / \kappa_T$ {\it must} always appear together. As I demonstrate below, this has fundamental consequences for the dynamics of  fluid instabilities in stably stratified, low P\'eclet number regions of stars, notably shear instabilities, and double-diffusive instabilities. 

\section{Stratified shear instabilities} 
\label{sec:stratshear}

\begin{figure}[h]
\includegraphics[width=\textwidth]{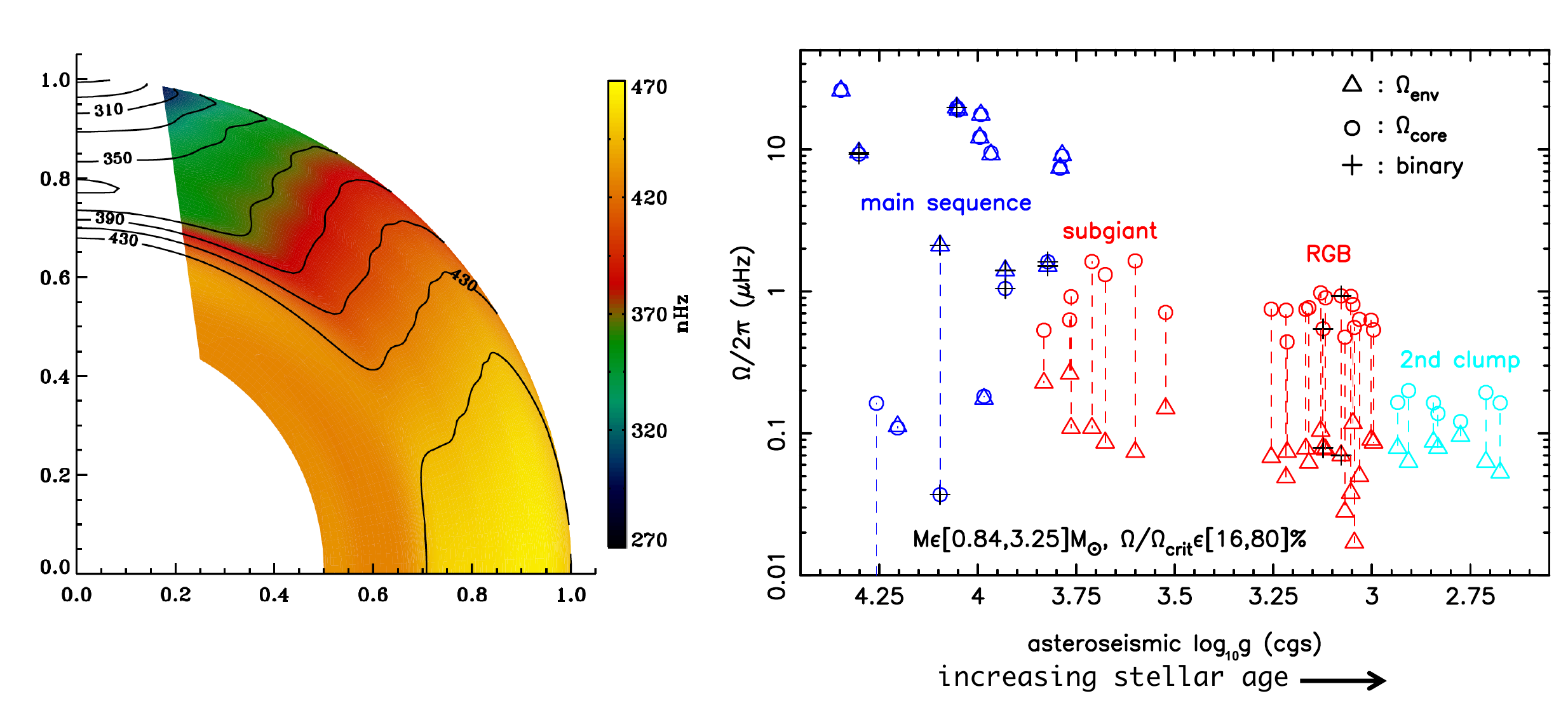}
\caption{Left: Angular rotation rate in the solar interior, in nHz, measured using helioseismology. Figure adapted from \citet{LarsonSchou2018}. Right: Envelope and core rotation rates of stars, measured using asteroseismology, as a function of surface gravity, which is a proxy for stellar age. Figure adapted from \citet{Aertsal2019}. }
\label{fig:rotprof}
\end{figure}

Shear is as omnipresent in stars as it is in the Earth's oceans and atmosphere. It exists on a huge range of scales and can take many forms, depending on its source. On the largest scales, the source of shear (in radiative zones) is almost always the star's differential rotation, which in turn comes from angular momentum conservation and transport by large-scale flows. It could also be driven by the thermal wind (i.e. horizontal gradients of temperature driving shear along the star's rotation axis), by tidal torques (if the star has a companion) or by magnetic torques. On intermediate scales, shear can be driven by meridional flows or large-scale internal waves. Until the 1980s, direct observational evidence for stellar shear was scarce; instead, its presence was generally inferred from stellar evolution models, which predict the development of substantial radial shear from angular momentum conservation as the star's core and envelope differentially expand and shrink. That has dramatically changed since the advent of helioseismology and more recently, asteroseismology, which now allow us to measure (or at least, estimate) the internal rotation profile of the Sun \citep{Thompson-etal96} and quite a few other stars (see the review by \citet{Aertsal2019}), see Figure \ref{fig:rotprof}. From these measurements, we know that stars exhibit both radial and latitudinal rotational shear \citep{Charbonneaual99}. In what follows, I shall therefore address both the effects of vertical and horizontal shear, ignoring for now the effects of rotation and magnetic fields  -- even though these are likely quite crucial to a complete understanding of the dynamics of stellar shear instabilities (see Section \ref{sec:magrot} for a short discussion of their effects). 

\subsection{Vertical shear}
\label{sec:vertshearintro}

\subsubsection{Context}

Vertical shear instabilities are an important source of diapycnal mixing in stratified geophysical flows, and have been studied in this context for over a hundred years. As laid out by \citet{Richardson1920}, the excitation and sustainability of turbulent perturbations to a stratified shear flow simply depends on the local energetics. If the kinetic energy transferred from the mean flow to the perturbations during a mixing event is larger than their potential energy cost, then the turbulence can be maintained. Otherwise, the perturbations must eventually decay. This is controlled by the local gradient Richardson number 
\begin{equation}
J = \frac{N^2}{S^2}, 
\end{equation}
where $N$ is the buoyancy frequency associated with the local total vertical potential temperature gradient $\beta_T + d\tilde T/dz$, and $S$ is the absolute value of the local vertical shearing rate. \citet{Richardson1920} argued that the quantity $J$ must drop below a certain threshold of order unity for turbulent perturbations to be maintained. While his argument was based on energetics, and can therefore be viewed as a statement on the nonlinear stability of the flow, \citet{Miles61} and \citet{Howard61} later formally proved that a necessary condition for linear instability of adiabatic perturbations in a unidirectional stratified shear flow is that $J$ should drop below 1/4 somewhere in the fluid. Both approaches thus point to the fact that one needs
\begin{equation}
J < J_c \mbox{  where  }  J_c  \sim O(1)
\end{equation}
 for shear instability. It is important to note, however, that the adiabaticity of the perturbations is a key assumption of both Richardson's nonlinear argument, and Miles and Howard's linear argument. As demonstrated by \citet{Townsend58}, radiative losses can relax this criterion, by reducing the relative buoyancy of the perturbations compared with the background, and therefore reducing their potential energy cost. More precisely, he showed that  provided the thermal adjustment timescale of the perturbations to the background is fast enough (i.e. provided $S t_{therm} \ll 1$ where $t_{therm}$ is the cooling / heating timescale of the perturbations), then the new criterion for instability should be
\begin{equation}
J  <  O\left( \frac{1}{S t_{therm}} \right) , 
\label{eq:RichardsonTownsend} 
\end{equation}
instead of $J < O(1)$. This Richardson-Townsend criterion thus allows for instability for  $J \gg1$ provided $t_{therm} \ll S^{-1}$. 

Meanwhile, and until the 1970s, vertical shear instabilities in stars had been given very little attention, presumably for two reasons. First, since shear on the largest scales is usually due to the star's differential rotation, the latter can hardly be ignored. When accounted for, rotation can either stabilize or destabilize the large-scale shear, while driving a variety of centrifugal instabilities depending on the direction of the angular momentum gradient \citep{Rayleigh1917,solberg1936mouvement,hoiland1941stability,GoldreichSchubert1967}. As such, much of the focus of research in those days was on centrifugally-driven instabilities, rather than on pure shear instabilities. Second, the typical values of the Richardson number derived from stellar evolution calculations are usually exceedingly large, ranging from $10^4$ and up. This is not entirely surprising: very roughly, if the shear is due to differential rotation then $S \sim \gamma \Omega_\star$ where $\gamma$ is a small number (otherwise, the star would have counter-rotating regions, which is rather unlikely). Then, assuming $N \simeq \bar N$ (recall that $\bar N$ characterizes the mean stratification), 
\begin{equation}
J \simeq  \frac{\bar N^2}{S^2} \sim \frac{g}{\bar \rho} \frac{|\partial \bar \rho/\partial r|}{\gamma^2 \Omega_\star^2} \sim  \frac{g}{ \gamma^2 r \Omega_\star^2} \sim  \frac{v^2_{esc}}{ \gamma^2 v^2_{rot}} , 
\end{equation}
where $\bar \rho(r)$ is the background radial density profile, $v_{esc} = \sqrt{2gr}$ is the gravitational escape velocity at a radius $r$, and  $v_{rot} = r\Omega_\star$ is the linear velocity associated with the star's rotation at the same position. We need $v_{rot} \ll v_{esc}$ for the star to be gravitationally bound, and since $\gamma$ must be small, $J$ is always very large. With the standard Richardson criterion in mind, it was therefore thought that shear instabilities would not be relevant in stellar interiors. 

\subsubsection{Zahn's stability criterion for diffusive shear instabilities}

This perception changed, however, primarily thanks to Jean-Paul Zahn. Zahn was an astrophysicist with broad interests in stellar hydrodynamics, who worked in New York with Spiegel in the late 1960s before moving back to France. While Spiegel's primary interest was convection, Zahn was more interested in the dynamics of stellar radiative zones, including small-scale turbulent mixing as well as transport by large-scale flows. Following Townsend's work on the effect of radiative losses on stratified shear flows in the atmosphere, \citet{SpiegelZahn1970} proposed that similar processes may be relevant in stellar interiors. \citet{Zahn1974} (see also \cite{Dudis1974,Jones1977}) quantified this idea by arguing that since stellar interiors are optically thick, the thermal adjustment timescale $t_{therm}$ should be related to the thermal diffusivity $\kappa_T$ and the characteristic size $l$ of perturbations as 
\begin{equation}
t_{therm} = \frac{l^2}{\kappa_T}.
\end{equation}
Using this in (\ref{eq:RichardsonTownsend}), Zahn obtained 
\begin{equation}
J  <  O\left( \frac{\kappa_T}{S l^2}\right)  \rightarrow J Pe_l < O(1)  \mbox{   provided } Pe_l \ll 1, 
\end{equation}
where $Pe_l = S l^2 / \kappa_T$ is the P\'eclet number based on the local shear and the eddy scale $l$. Naively, one may therefore argue that by taking $l$ to be as small as possible, $Pe_l \ll 1$ and $J Pe_l < O(1)$ can always be satisfied. This would suggest that {\it any} amount of shear could become unstable. However, this is not the case: Zahn further noted that perturbations cannot be so small as to become viscously controlled. He therefore also required that the Reynolds number based on the same eddy scale should be greater than a certain threshold for instability, which he estimated to be $O(1000)$ based on available laboratory experiments at the time. Mathematically, 
\begin{equation}
Re_l = \frac{Sl^2}{\nu} > Re_c \sim O(10^3).
\end{equation}
Combining the two yields the requirement that 
\begin{equation} 
\frac{J  Pe_l }{Re_l} < O(Re_c^{-1})  \rightarrow JPr < (JPr)_c, 
\label{eq:Zahncrit}
\end{equation}
 where $(JPr)_c \sim O(10^{-3})$. We therefore see that, according to Zahn's criterion \citep{Zahn1974}, stratified shear instabilities can exist for fairly large $J$ in the low Prandtl number environments of stellar interiors. Taking $Pr \sim 10^{-6}$, for instance, instabilities should be present up to $J \sim 10^3$.

It is important to note, however, that Zahn's argument does not derive from any rigorous linear stability analysis; instead, it can be viewed on a par with Richardson's original discussion of the energetics of stratified turbulence. As such, it is particularly important to check its validity. Its form, however, is not unexpected when viewed from the perspective of the low P\'eclet number asymptotic equations \citep{Lignieres1999}, which are relevant here since Zahn explicitly assumes that $Pe_l \ll 1$. Indeed, looking at equation (\ref{eq:LPNmomentum}), we see that the only relevant dimensional parameters and groups of parameters are $\nu$ and the combination $\bar N^2 / \kappa_T$, together with the amplitude $S$ of the background shear\footnote{Since the eddies are supposedly small, they can only know about the background shear rather than other large-scale properties of the flow}. The only way to create a non-dimensional quantity involving the Richardson number $N^2/S^2$ (noting that $N \simeq \bar N$ in the low P\'eclet number limit) using the available dimensional parameters is $(\bar N^2/\kappa_T) ( S^2 / \nu)^{-1}$, which is indeed $J Pr$. 

\begin{figure}[h]
\includegraphics[width=\textwidth]{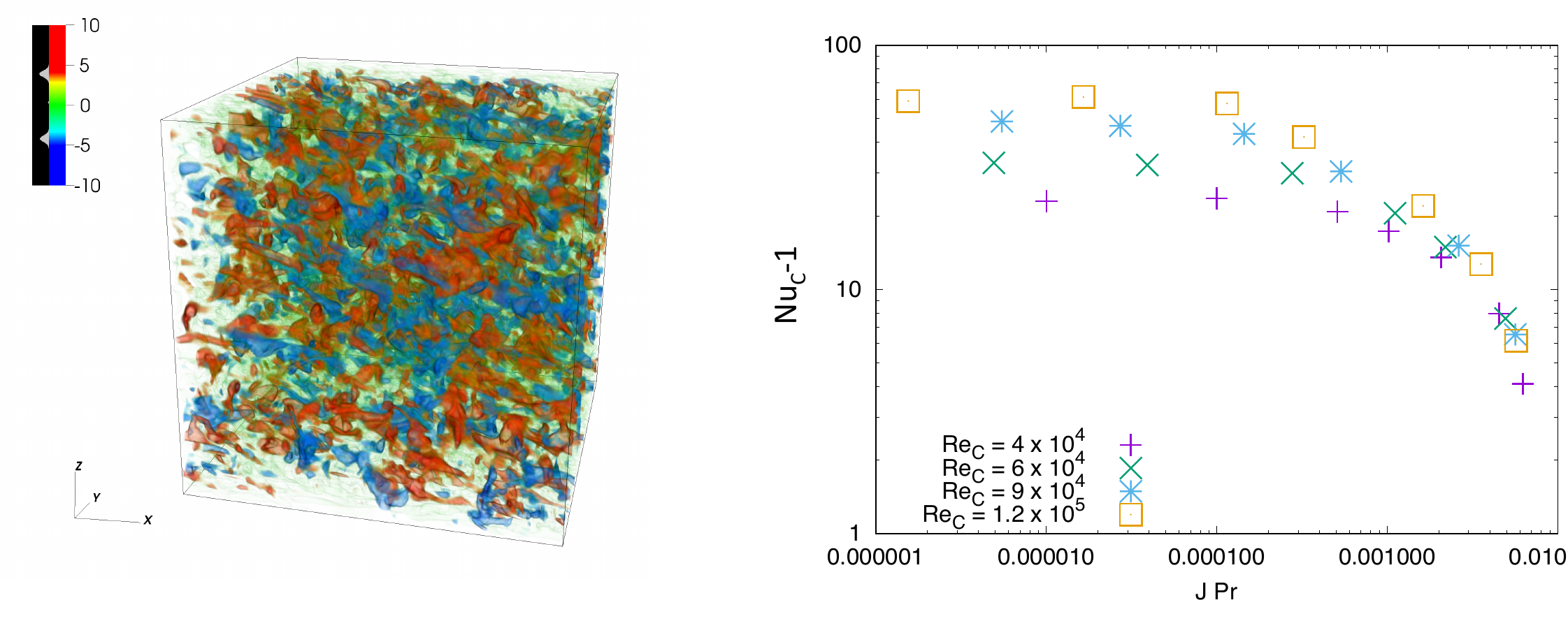}
\caption{Left: Volume-rendered snapshot of the nondimensional vertical velocity field in a simulation of stratified plane Couette flow using the LPN approximation. Adapted from Figure 2 of \citet{Garaudal17}. Right: Measurements of the nondimensional turbulent diffusivity (expressed as a Nusselt number) for a passive scalar in the same simulations, as a function of $JPr$ measured in the bulk of the shear layer. Note how the turbulence disappears as $JPr \rightarrow (JPr)_c \simeq 0.007$. Figure created from the data presented in \citet{Garaudal17}. }
\label{fig:VertStratShear}
\end{figure}

It took 40 years, however, for technological advances in high-performance computing to enable the scientific community to test Zahn's theory. Even today, computational limitations (especially in 3D, which is necessary for a reliable test) force us to use simulation parameters that are very far from those of stellar interiors (where $Pr \ll 1$ and $Re \gg 1$). The first attempt at testing Zahn's theory using DNS was presented by \citet{PratLignieres13} (see also \cite{PratLignieres14,Pratal2016}) for the case of a homogeneous shear flow. Later, we attacked the same problem using different model setups: \citet{GaraudKulen16} studied the stability of a stratified Kolmogorov flow (i.e. sinusoidal flow) driven by a body force, while \citet{Garaudal17} considered a stratified plane Couette flow (see Figure \ref{fig:VertStratShear}). Remarkably, {\it} all of these studies came to the same conclusion: in the limit where the turbulent P\'eclet number is small, turbulence cannot be sustained if $JPr > 0.007$ everywhere in the flow. The same result was obtained using both the standard equations at low P\'eclet number, and the low P\'eclet number (LPN) asymptotic equations (\ref{eq:LPNmomentum}). 
The agreement in the stability threshold identified in these very different kinds of DNS is quite remarkable, and fully validates Zahn's criterion \citep{Zahn1974} for stratified shear instabilities {\it at low P\'eclet number}. However, whether this criterion still applies for {\it all} low Prandtl number flows, including those that have a large {\it outer scale} P\'eclet number $Pe$ (where $Pe = \bar SL^2/\kappa_T$, with $L$ a measure of the width of the shear layer and $\bar S$ is its mean shear) as Zahn originally intended, remains an open question.  

 A completely independent formal approach to the question was also proposed in \cite{Garaudal15} (based on the Woods Hole GFD summer program project of Tobias Bischoff, who was a fellow in 2013), where we used energy stability arguments  to demonstrate that for sufficiently large Reynolds number, {\it any} perturbation to a low P\'eclet number Kolmogorov flow must decay unless $JPr < (JPr)_E$, where $(JPr)_E \sim  O(1)$ is a constant. This criterion has the same form as Zahn's criterion (\ref{eq:Zahncrit}), albeit with a constant that is $O(1)$ on the right-hand side instead of $\sim 0.007$. Note that the energy stability argument begins with the LPN approximation (\ref{eq:LPNmomentum}), rather than the standard equations (\ref{eq:momentum})-(\ref{eq:heat}). Indeed, using the standard equations would only provide an energy stability criterion that is independent of the Richardson number. This is because it is always possible to initialize the flow with perturbations that locally reduce the stratification to the point where $J$ drops below 1, which then permits the development of shear instabilities. By contrast, the LPN approximation does not allow substantial modification of the background stratification (see Section \ref{sec:LPNapprox}) and instead imposes a very tight constraint between the vertical velocity field and the temperature fluctuations. This constraint restricts the parameter space of allowable perturbations, and therefore provides an energy stability bound that depends on the stratification.

\subsubsection{Zahn's model for vertical mixing by vertical shear instabilities} 
 \label{sec:vertshearmix}

Later on, Zahn became interested in quantifying the rate of mixing by shear-induced turbulence in stellar interiors \citep{Zahn92}. In this now classical 1992 paper, he put forward a simple model for the turbulent mixing coefficient $D$ for vertical shear instabilities (this derivation could equally apply to a turbulent compositional diffusivity, or a turbulent viscosity). 
To do so, he first noted that, from a dimensional perspective, 
\begin{equation}
D \propto S l^2, 
\label{eq:Dsl2}
\end{equation}
where $l$ is the typical size of energy-bearing eddies in the turbulent flow and $S$ is the local shearing rate (recall that $S$ is positive by definition). As we saw earlier, perturbations on a scale $l$ can grow provided $J Pe_l < O(1)$, so Zahn suggested one should use the largest possible value of $l$ for which this statement holds. This quantity is now known as the Zahn scale and satisfies 
\begin{equation}
J \frac{ S l_Z^2}{\kappa_T}   \sim O(1) \rightarrow l_Z  \sim \sqrt{\frac{\kappa_TS}{\bar N^2} } .
\end{equation}
 It is not difficult to see that this is in fact the {\it only} lengthscale that can be constructed from the dimensional groups $\bar N^2 / \kappa_T$ and $S$ (ignoring viscosity), so the form of $l_Z$ should not come as a surprise. It is, however, strikingly different from the lengthscale often associated with high Reynolds number stratified turbulence in geophysical flows, namely $U/\bar N$ \citep{BillantChomaz2001,Brethouweral2007}; see more on this later. 
Using $l_Z$ in (\ref{eq:Dsl2}), we obtain
\begin{equation}
D \propto S l_Z^2 = C_Z \frac{\kappa_T}{J},
\label{eq:ZahnDv}
\end{equation}
where $C_Z$ is a constant of order unity. 
This estimate should hold as long as $l_Z$ is much smaller than other available characteristic scales of the fluid, say $L$, (such as, e.g. the local density scaleheight, or the scaleheight of the background shear, etc.), and as long as
the flow is indeed unstable, that is when $JPr < (JPr)_c$ (see equation \ref{eq:Zahncrit}). This stability criterion, incidentally, can now easily be shown to be equivalent to $l_Z > l_\nu$, where $\l_\nu$ is the scale below which viscous effects become important, i.e. the scale for which $Re_\nu = Sl_\nu^2 / \nu = Re_c$ (see equation \ref{eq:Zahncrit}).  As such, (\ref{eq:ZahnDv}) should be valid as long as there is a separation of scales satisfying
\begin{equation}
l_\nu \ll l_Z \ll L.
\label{eq:lzcond}
\end{equation}

There have been several attempts at comparing Zahn's turbulent mixing prescription with DNS, notably  by \citet{PratLignieres14}, \citet{Pratal2016}, and \citet{Garaudal17}. Most of these focussed on cases where the outer scale P\'eclet number is small, ensuring that the turbulence is in the low P\'eclet number regime. The comparison between theory and experiments is numerically challenging, because the condition (\ref{eq:lzcond}) requires a very large dynamical range to be satisfied. In fact, it is not entirely clear that any of the DNS presented to date actually reach such a clear separation of scales, but at least those at the highest available Reynolds numbers have $L > 20 l_\nu$, and therefore approach it. For sufficiently high Reynolds numbers, the turbulent diffusivities measured by \citet{Pratal2016}, and \citet{Garaudal17} are again remarkably consistent, despite using very different model setups. For all simulations at low P\'eclet number where (\ref{eq:lzcond}) is satisfied, (\ref{eq:ZahnDv}) holds with $C_Z \sim O(0.1)$. This is a tentative result, however, that will need to be confirmed with simulations at much higher Reynolds number, of $O(10^6)$ at least, see \cite{Garaudal17}. 
 It is also interesting to note that in all of the DNS performed in which both are measured, the turbulent diffusivity for a passive scalar and the turbulent viscosity are consistent within 10 or 20 percent  -- in other words, the turbulent Schmidt number is close to one. As such, when valid, Zahn's model (\ref{eq:ZahnDv}) should provide a good order magnitude estimate for both the turbulent diffusivity {\it and} the turbulent viscosity in stars (assuming rotation and magnetic fields can be ignored, which is not necessarily the case, see Section \ref{sec:magrot}). 

\subsubsection{Stellar implications} 

In summary, there seems to be a wealth of evidence in favor of Zahn's stability criterion (\ref{eq:Zahncrit}) for low P\'eclet number vertical shear flows in stars. There is also tentative evidence that his model for turbulent mixing by vertical shear instabilities (\ref{eq:ZahnDv}) may apply for low P\'eclet number flows, although simulations with a much larger dynamical range will be required to establish whether this result is robust. However, it remains to be determined whether the stability criterion and the turbulent mixing prescription are valid more generally for all shear flows at low Prandtl number as originally proposed in \citet{Zahn1974} and \citet{Zahn92}.  

From a stellar perspective, this current restriction on the applicability of Zahn's models to low P\'eclet number shear layers is unfortunate. Indeed, it is not possible to compute the turbulent P\'eclet number without knowing the actual eddy size, and the latter cannot be observed. As a result, we do not know, simply from observations, whether the flow satisfies the LPN approximation or not (which is a necessary condition for Zahn's models to apply). At best, we can compute the outer scale P\'eclet number $Pe$ based on the observed shear and use it as an upper limit to the turbulent P\'eclet number. But as discussed by \citet{GaraudKulen16}, with reasonable assumptions on the amplitude of the shear, $Pe$ is likely very large in stars (see equation (\ref{eq:tachopars}) below for a quantitative estimate), except perhaps in the outer layers of the most massive stars, where $\kappa_T$ can be in excess of $10^{15}$cm$^2$/s. 
 Deep in the interior of solar-type or intermediate mass stars, for instance, $Pe$ is of the order of $10^5-10^8$ so Zahn's models cannot be safely applied (yet). In order to do so, one would need to demonstrate that small-scale perturbations to the large-scale shear can always be excited nonlinearly, which would then result in a much lower turbulent P\'eclet number. This is an open question that remains to be addressed, and will require simulations in much wider computational domains and at much higher Reynolds number than presently available. 

Even assuming that Zahn's estimate (\ref{eq:ZahnDv}) holds generally for any low $Pr$ flow, the general conclusion is that vertical shear instabilities are {\it not} a particularly important source of turbulent mixing in stars, for two reasons. First, combining the criterion $JPr < 0.007$ with $Pr \sim 10^{-6}$ implies that $J$ cannot be larger than about $10^4$ for instability to occur. While some stellar shear layers do indeed satisfy this (e.g. the solar tachocline has $\bar N^2 \sim 10^{-6}$s$^{-2}$ and $\bar S \sim 10^{-5}$s$^{-1}$, so on average $J \sim 10^4$), this is uncommon for the reasons discussed in Section \ref{sec:vertshearintro}. Secondly, even if the flow is unstable according to (\ref{eq:Zahncrit}), we have 
\begin{equation} 
D \simeq \frac{C_Z}{(JPr)_c} \frac{(JPr)_c}{JPr}   \nu \simeq 10 \frac{(JPr)_c}{JPr} \nu
\end{equation} 
using the estimated numerical values of $C_Z \sim O(0.1)$ and $(JPr)_c \sim O(0.01)$. As such, unless $JPr \ll (JPr)_c$, the turbulent viscosity is predicted to be only one or two orders of magnitude larger than its microscopic counterpart. The same is true for the turbulent diffusivity of a scalar field, since the microscopic compositional diffusivity is usually of the same order as $\nu$. As such, we are forced to conclude that stratified vertical shear instabilities in low Prandtl number stellar fluids are not a substantial source of turbulent transport for momentum or composition. 

\subsection{Horizontal shear}

\subsubsection{Context} 

The effect of horizontal shear on mixing in stratified fluids is much less straightforward than that of vertical shear, but the two are not unrelated. Consider for instance a unidirectional, vertically invariant, horizontal shear flow of the form $U(y){\bf e}_x$, where ${\bf e}_x$ is the streamwise direction, and $y$ is the spanwise direction. A two-dimensional (2D) perturbation to this flow, which takes the form of horizontal motions only, is unaffected by stratification. As such, this 2D perturbation will be linearly unstable provided $U(y)$ satisfies the standard criteria for 2D, unstratified shear instabilities (see, e.g. \cite{DrazinReid}). One may therefore argue that horizontal shear instabilities are much easier to trigger than vertical shear instabilities. However, the same 2D motions cannot cause any vertical mixing, so the latter must result from 3D perturbations, which are, by contrast, directly affected by the stratification. The goal is therefore to understand how these 3D motions are driven, what form they take, and how much diapycnal mixing they can cause.  

On Earth, evidence for vertical mixing in (mostly) horizontal shear flows is quite clear from both laboratory experiments \citep{Parkal1994,HolfordLinden1999,Oglethorpeal2013,Thorpe2016} and numerical experiments \citep{Brethouweral2007,Lucasal2017}. In all cases, small-scale vertical shear is an essential component of the process, and can appear for a variety of reasons, either because the driving mechanism for the horizontal motions is not vertically invariant, or because 3D modes of instability are also excited \cite{BillantChomaz2000,Oglethorpeal2013,Lucasal2017}. In itself, this vertical shear might not be sufficient to trigger vertical shear instabilities, because the gradient Richardson number constructed using the shear and the mean background stratification remains very large. However, by virtue of non-monotonic buoyancy flux laws \citep{Phillips1972,ballsy}, the stratification crucially rearranges itself to contain alternating  regions that are more weakly and more strong stratified, respectively -- as a set of layers and interfaces. The weaker stratification within the layers locally decreases the gradient Richardson number below unity, allowing the flow to  become turbulent. The turbulence, in turn, mixes the layers and maintains the interfaces so the process is essentially self-sustaining \cite{Caulfield2021}. 

It has been shown both experimentally, theoretically and numerically, that in the limit where viscosity is negligible, the thickness of these layers scales as $U_h/\bar N$, where $U_h$ is the r.m.s. amplitude of the horizontal flow (see, e.g. \cite{Parkal1994,HolfordLinden1999,BillantChomaz2001,Brethouweral2007,Oglethorpeal2013}). This scaling is again not surprising: in geophysical environments where viscosity is negligible, so is the thermal diffusivity since $Pr \sim O(1)$ or larger, and the only available dimensional parameters of the system are the characteristic scale of the horizontal flow $L$, the amplitude of this flow $U_h$ (or, the horizontal shear $S_h \sim U_h/L$, depending on the model setup), as well as the mean stratification $\bar N$. The only lengthscale that can be constructed from $U_h$ and $\bar N$ independently of $L$ is $U_h/\bar N$. 

From the discussions presented in Section \ref{sec:LPNapprox}, however, it is quite clear that a pathway to turbulence involving the formation of layers and interfaces is prohibited at low P\'eclet number, since the background stratification cannot be modified in that limit. Furthermore, since the relevant dimensional parameter is $\bar N^2 / \kappa_T$ (rather than $\bar N$ and $\kappa_T$ separately), the quantity $U_h/\bar N$ cannot be the appropriate lengthscale for low P\'eclet number flows, as we already found in the case of vertical shear (see Section \ref{sec:vertshearmix}). In other words, the dynamics of horizontal shear instabilities in stars {\it must} be quite different from those on Earth. 

The first model for turbulent mixing induced by horizontal shear instabilities in stellar interiors was proposed by Zahn, in the same 1992 paper that introduced the mixing coefficient for vertical shear instabilities \citep{Zahn92}. Zahn argued that the presence of horizontal shear would drive primarily 2D motions, which rapidly decouple in the vertical direction owing to the very low plasma viscosity. These now layerwise horizontal motions become unstable to diffusive vertical shear instabilities when their characteristic vertical scale $l_v$ drops below the Zahn scale, i.e. when $l_v = \sqrt{\frac{\kappa_TS}{\bar N^2} }$, where here $S = U_h / l_v $. Solving for $l_v$ yields the new scaling
\begin{equation}
l_v =  \left( \frac{\kappa_T U_h}{\bar N^2} \right)^{1/3} , 
\end{equation}
which is (again) the only lengthscale that can be constructed from the dimensional quantities $U_h$ and $\bar N^2 / \kappa_T$ (see Section \ref{sec:LPNapprox}; here, the low P\'eclet number approximation is implicit since the vertical shear instabilities are assumed to be diffusive). Zahn further assumed that the amplitude of the horizontal motions can be obtained from the viscous dissipation rate $\epsilon$ using the usual Kolmogorov scaling $\epsilon \propto  U_h^3/l_v$, ultimately leading to the conclusion\footnote{Note that Zahn never explicitly wrote $l_v$ and $U_h$ as such, but it can be inferred from his calculation.} that 
\begin{equation}
l_v  =   \left( \frac{\kappa_T \epsilon^{1/3}}{\bar N^2} \right)^{3/8} \mbox{   and  } U_h =  \left( \frac{\kappa_T \epsilon^{3}}{\bar N^2} \right)^{1/8}  . 
\label{eq:horizahn}
\end{equation}
\citet{Lignieres2020} recently provided an alternative explanation for this scaling, which may feel more familiar to geophysical fluid dynamicists. In high P\'eclet number flows, he recalls, the effects of stratification begin to affect the turbulence above the Ozmidov scale $l_O$, which is the scale at which the eddy turnover timescale $l/u(l)$ equal the buoyancy timescale $\bar N^{-1}$. Below the Ozmidov scale, the turbulence satisfies the usual Kolmogorov scaling relating the dissipation rate to the turbulent spectrum $\epsilon \propto u(l)^3 / l$. Together, 
\begin{equation}
\frac{l_O}{u(l_O)} = \bar N^{-1} \mbox{ and  } \epsilon = \frac{u^3(l_O)}{l_O} \Rightarrow l_O = \left( \frac{\epsilon}{\bar N^3} \right)^{1/2} .
\end{equation}
For low P\'eclet number flows, on the other hand, Ligni\`eres argues that this argument needs to be modified to account for the fact that the buoyancy timescale is no longer $\bar N^{-1}$, but (by dimensional analysis) $\kappa_T/\bar N^2 l^2$. Equating this with the eddy turnover timescale, with the same constraint from the energy dissipation rate, he defines a modified Ozmidov scale $l_{OM}$ as
\begin{equation}
\frac{l_{OM}}{u(l_{OM})} = \frac{ \kappa_T}{\bar N^2 l_{OM}^2}  \mbox{ and  } \epsilon = \frac{u^3(l_{OM})}{l_{OM}} \Rightarrow l_{OM} = \left( \frac{\kappa_T \epsilon^{1/3}}{\bar N^2} \right)^{3/8} . 
\end{equation}
This recovers (\ref{eq:horizahn}), as mentioned earlier, and provides insight into the balance of timescales required in deriving $l_v$.

\citet{Zahn92} then used the derived lengthscale and velocity amplitude to compute a vertical turbulent mixing coefficient, which is 
\begin{equation}
D \propto l_v U_h \propto \left( \frac{\kappa_T \epsilon}{\bar N^2} \right)^{1/2} . 
\label{eq:ZahnDh}
\end{equation}
This predicted scaling is interesting, because it only depends on the stratification as $\bar N^{-1}$, compared with the mixing coefficient for vertical shear instabilities, which scales as $\bar N^{-2}$. As such, it is possible that horizontal shear instabilities may actually provide a more efficient source of mixing in stars than vertical shear instabilities in the limit of strong stratification. 

\subsubsection{Direct Numerical simulations of horizontal shear instabilities at low Prandtl number}

Having worked on the problem of vertical shear instabilities for a few years, I decided to tackle the more complicated problem of horizontal shear instabilities in 2018. By chance, Colm-cille Caulfield and I were both planning to attend the GFD program that summer. Caulfield has done extensive work on stratified turbulence with application to the oceanographic context, and had recently studied stratified horizontal Kolmogorov flows at $Pr \sim O(1)$ \cite{Lucasal2017}. Together with GFD fellow Laura Cope, we extended that work to the low $Pr$ limit, which paved the way to a comprehensive analysis of Zahn's model for turbulent mixing by horizontal shear flows in stars. As we discovered \cite{Copeal20,Garaud20} the story is substantially more complicated than Zahn foresaw. 

As a natural continuation of prior work \cite{GaraudKulen16,Lucasal2017}, we studied the dynamics of horizontal Kolmogorov flows in a vertically-stratified fluid. The governing equations are given by (\ref{eq:momentum})-(\ref{eq:heat}), with the addition of a body force of the form ${\bf F} = F_0 \sin(k_s y) {\bf e}_x$ to drive the horizontal shear. Using a non-dimensionalization based on the outer scale of the flow $k_s^{-1}$, the predicted amplitude of the horizontal flow $U_h = \sqrt{F_0 / \rho_m k_s}$, and the temperature scale $k_s^{-1} \beta_T$, we arrive at equations very similar to (\ref{eq:lignieresmomentum})-(\ref{eq:lignieresheat}), with $B = \bar N^2 / U_h^2 k_s^2$, $Re_t$ replaced by an outer scale Reynolds number $Re = U_h /\nu k_s$ and $Pe_t$ replaced by an outer scale P\'eclet number $Pe = U_h /\kappa_T k_s$. Note that the quantity $U_h$ turns out to be a good estimate for the horizontal rms velocity, hence the choice to keep the same notation as in the previous section. For typical parameters of the interiors of main sequence stars of a few solar masses or less, $Re$, $Pe$ and $B$ are all much larger than one, with $Pe = Pr Re \ll Re$ \cite{Garaud20}. For the outer layers of very high mass stars, by contrast, $Re \gg 1$ while $Pe < 1$ \citep{GaraudKulen16}. These estimates show that both $Pe \gg 1$ and $Pe < 1$ regimes are relevant in stellar contexts. 

\begin{figure}[h]
\includegraphics[width=\textwidth]{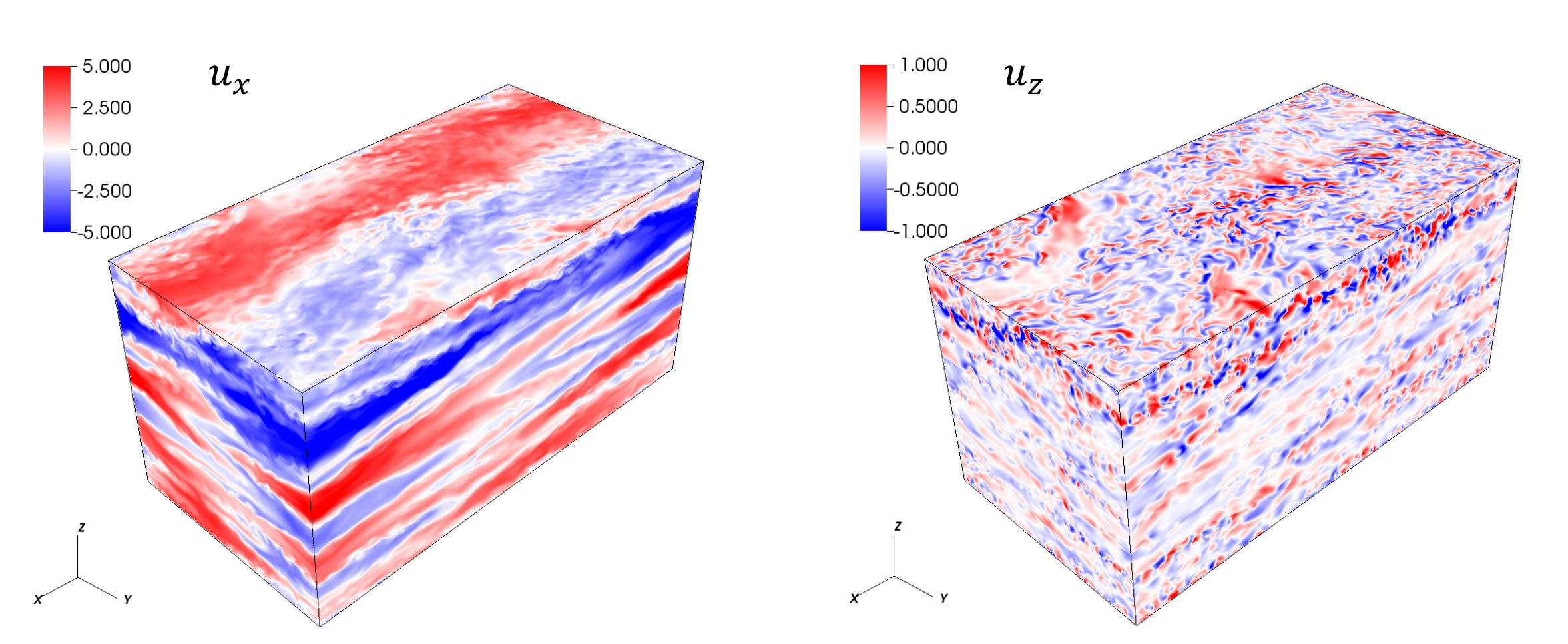}
\caption{Snapshots of the horizontal (left) and vertical (right) velocity field in a simulation of horizontal shear flow at $Re = 600$, $Pe = 0.1$, $B = 6000$, discussed in \citet{Copeal20}. The meanders of the horizontal flows are visible on the left, and generate substantial shear, that drives vertical shear instabilities on small scales (right).}
\label{fig:SnapshotsHS}
\end{figure}

As predicted by \citet{Zahn92}, in \citet{Copeal20} we found that the mean horizontal flow rapidly becomes unstable to quasi-2D perturbations, that take the form of vertically modulated horizontal meanders. These decoupled meanders induce some vertical shear, that can become unstable and cause vertical mixing, depending on the stratification. For weakly stratified flows, the turbulence behaves as if temperature was a passive scalar, and is unaffected by stratification. For more strongly stratified flows, both vertical eddy size and rms vertical velocity are reduced by the stratification, while the horizontal rms velocities remain essentially unaffected (see Figure \ref{fig:SnapshotsHS}). For even more strongly stratified flows, the vertical shear between the meanders is progressively stabilized, first intermittently, and then entirely. Beyond this qualitative picture, however, the results are sensitively dependent on the {\it emergent} turbulent P\'eclet number of the {\it vertical} fluid motions, given by $Pe_t = w_{rms} l_v / \kappa_T$, where $w_{rms}$ is the vertical rms velocity and $l_v$ is the vertical eddy scale. This is irrespective of $Pe$, although we always have $Pe_t = (w_{rms}/U_h) (l_v k_s) Pe < Pe$. 

In the limit where $Pe_t \gg 1$ and the flow is strongly stratified ($B>1$),
\begin{equation}
l_v \propto B^{-1/3}  k_s^{-1} \mbox{   and   } w_{rms} \propto  B^{-1/3} U_h,
\label{eq:Garaudlv}
\end{equation}
independently of $Re$ or $Pe$, see \cite{Garaud20}. This would then imply a turbulent mixing coefficient 
\begin{equation}
D \propto l_v w_{rms} \sim B^{-2/3} k_s^{-1} U_h \propto \left( \frac{ \bar N^2}{U_h^{7/2} k_s^{1/2}  }   \right)^{-2/3}   ,
\end{equation}
so $D \propto \bar N^{-4/3}$, which decays a little faster than Zahn's prediction (\ref{eq:ZahnDh}), but not as fast as the mixing coefficient for vertical shear instabilities (\ref{eq:ZahnDv}). 
In addition, this kind of turbulence appears to be capable of inducing a non-negligible heat flux (compared with the diffusive flux), estimated to be \cite{Garaud20}
\begin{equation}
F_T \propto -  \frac{ U_h}{B k_s} \beta_T \propto  -  \frac{ U_h^3 k_s}{\alpha_T g } .
\end{equation}
Note that the heat flux is downward in stably stratified fluids, and usually has the opposite sign to the conducted (radiative) flux since the temperature gradient $T_{0z}$ is negative in stars. 

To my knowledge, these scalings are different from any other proposed to date, but naturally arise from a dominant balance between nonlinear terms and forcing in the horizontal component of the momentum equation, hydrostatics in the vertical direction, and a balance between the nonlinear terms ${\bf u} \cdot \nabla \tilde T$ and the background advection term $\beta_T w$ in the temperature equation (see \cite{Garaud20} for detail). The fact that the vertical eddy scale in (\ref{eq:Garaudlv}) is not given by $U_h / \bar N$, for instance, is particularly surprising since in this limit both Reynolds and P\'eclet numbers are relatively large, and one may expect geophysically-relevant scaling arguments such as those of \citet{Brethouweral2007} to apply. However, it is worth noting that the $U_h / \bar N$ scaling is formally only applicable when the buoyancy Reynolds number defined as $Re_b = \epsilon/ \nu \bar N^2$ is very large, and it is likely that the present simulations (for which $Re_b$ ranges from 3 to 20 in the regime of interest) are not in that limit yet. As such, these results remain to be confirmed in the future. 

From (\ref{eq:Garaudlv}), we see that the turbulent P\'eclet number $Pe_t = w_{rms} l_v / \kappa_T$ must decrease with decreasing $Pe$ or increasing $B$. The critical $Pe_t = 1$ transition occurs when $B \propto Pe^{3/2}$, irrespective of $Re$ (or the Prandtl number). 
 When $Pe_t \ll 1$, the flow dynamics are well represented by the LPN approximation, and depend only on $Re$ and the product $BPe$, for the reasons described in Section \ref{sec:LPNapprox}. Various dynamical regimes exist, 
including a low P\'eclet number stratified turbulence regime, an intermittent regime, and a viscous regime, each following a distinct set of scaling laws discussed by Cope et al.  \citep{Copeal20}. In the turbulent low P\'eclet number stratified regime, for instance, the vertical eddy size and vertical r.m.s. velocity scale as 
\begin{equation}
l_v \propto (BPe)^{-1/3} k_s^{-1}, \mbox{   and   } w_{rms} \propto (BPe)^{-1/6} U_h,
\end{equation}
leading to a turbulent mixing coefficient 
\begin{equation}
D \propto l_v w_{rms} \propto (BPe)^{-1/2} k_s^{-1} U_h \propto \left( \frac{ \bar N^2}{U_h^3 k_s \kappa_T  }\right)^{-1/2}   .
\end{equation}
These scalings can be explained from a dominant balance between nonlinear terms and forcing in the horizontal component of the momentum equation (as before), nonlinear terms and buoyancy force in the vertical component of the momentum equation, and finally, the LPN balance in the thermal energy equation. We find that they do recover Zahn's prediction (\ref{eq:ZahnDh}) assuming that the viscous dissipation $\epsilon$ is given by the Kolmogorov scaling $U_h^3 k_s$ (which remains to be determined). We therefore confirm that $D \propto \bar N^{-1}$ in this regime, suggesting that mixing can remain important even when the stratification is strong. 

For very large values of $B$, finally, the vertical shear gradually becomes more stable, and the volume fraction of the domain occupied by turbulent flow decreases \citep[see][for more detail]{Copeal20}. Instead, viscously dominated flow structures emerge, with a vertical scale proportional to $Re^{-1/2}$. Vertical mixing in that regime is essentially negligible. 

\subsubsection{Stellar implications} 

Using values of $\kappa_T$, $\bar N$, $U_h$ and $k_s$ that are somewhat typical\footnote{Aside from the mean horizontal flow velocity $U_h$, which can vary significantly, the typical values of $\kappa_T$ and $\bar N$ do not change too much over the interior of the star, except perhaps close to the edge of the convection zone where $\bar N \rightarrow 0$. These values do not change much with stellar mass either, in the range $\sim 0.5 M_\odot$ to $\sim 10M_\odot$; the value of $k_s$ is taken to be of the order of $2\pi$ over the stellar radius, which does not change too much either.} of the deep interiors of Main Sequence stars with masses around $1M_\odot$, we have
\begin{eqnarray}
Re = 10^{14} \left(\frac{U_h}{10^4 {\rm cm/s}} \right) \left(  \frac{k_s^{-1}}{10^{11} {\rm cm}} \right) \left( \frac{\nu}{10{\rm cm^2/s}}  \right)^{-1},  \nonumber \\
Pe = 10^{8}  \left(\frac{U_h}{10^4 {\rm cm/s}} \right) \left(  \frac{k_s^{-1}}{10^{11} {\rm cm}} \right) \left( \frac{\kappa_T}{10^7{\rm cm^2/s}}  \right)^{-1}, \nonumber \\
B = 10^{8}  \left(\frac{U_h}{10^4 {\rm cm/s}} \right)^{-2} \left(  \frac{k_s^{-1}}{10^{11} {\rm cm}} \right)^2  \left( \frac{\bar N}{10^{-3}{\rm s}^{-1} } \right)^2,
\label{eq:tachopars}
\end{eqnarray}
so that, in the high $Pe_t$ regime, 
\begin{equation}
D \propto 10^{29/3}   \left(\frac{U_h}{10^4 {\rm cm/s}} \right)^{7/3} \left(  \frac{k_s^{-1}}{10^{11} {\rm cm}} \right)^{-1/3}  \left( \frac{\bar N}{10^{-3}{\rm s}^{-1} } \right)^{-4/3}   {\rm cm^2/s} , 
\end{equation}
while in the low  $Pe_t$ regime
\begin{equation}
D \propto   10^{7}  \left(\frac{U_h}{10^4 {\rm cm/s}} \right)^{3/2}  \left( \frac{\bar N}{10^{-3}{\rm s}^{-1} } \right)^{-1}   \left( \frac{\kappa_T}{10^7{\rm cm^2/s}}  \right)^{1/2} \left(  \frac{k_s^{-1}}{10^{11} {\rm cm}} \right)^{-1/2}  {\rm cm^2/s} .
\end{equation}
We see that in both cases, unless $U_h \ll 10^4$cm/s, $D$ is significantly larger than the microscopic viscosity $\nu \simeq 10$cm$^2$/s (and the microscopic diffusivity for a chemical tracer, which is of the same order of magnitude), suggesting that horizontal shear instabilities in either regime could be a substantial source of vertical mixing in stars, just as it is in the ocean and in the atmosphere on Earth. Of course, these results do not yet account for the effects of rotation or magnetic fields which are expected to be important in stars (see Section \ref{sec:magrot} and, e.g. \cite{Zahn92,Garaud20} for a discussion of these effects). 

In summary, we saw that a small Prandtl number is very favorable to the development of shear instabilities in strongly stratified fluids because it allows for the existence of a new regime which is both almost inviscid {\it yet also} thermally diffusive, i.e. such that $Pe_l = Pr Re_l  \ll 1 \ll Re_l$ on some lengthscales $l$. In this regime, thermal diffusion reduces the stabilizing effects of stratification and the instability can thrive, albeit at small scales. 
This conclusion, as we shall now see, applies to double-diffusive instabilities as well.

\section{Double-diffusive instabilities}
\label{sec:DDC}

A third group of instabilities that are of interest to both astrophysical and  geophysical communities are double-diffusive instabilities. These were discovered by \citet{Stommelal1956} and \citet{stern1960sfa}, in the context of their research in physical oceanography. \citet{stern1960sfa} realized that because the density of seawater depends on both temperature and salinity, which diffuse at very different rates, two new kinds of instabilities exist that can destabilize a statically stable density stratification (even more were discovered later, e.g. \cite{Holyer1983,Radko2016}). More generally, the same is true of any fluid whose density depends on multiple components that diffuse at different rates. The simplest model setup that supports double-diffusive instabilities is that of an unbounded fluid with a background potential temperature gradient $\beta_T$ and a background salinity (or any slow diffusing compositional field $C$) gradient $\beta_C$ (see \cite{baines1969,radko2013double}). The dimensional equations governing the fluid are as in (\ref{eq:momentum})-(\ref{eq:heat}), with the added contribution of the compositional perturbations $\tilde C$ to the buoyancy field, and a second advection diffusion equation: 
\begin{eqnarray}
\frac{\partial{\bf u}}{\partial t} + {\bf u}\cdot\nabla {\bf u} & = & - \frac{1}{\rho_m} \nabla \tilde{p} + (\alpha_T \tilde{T}  - \alpha_C \tilde{C} ) g {\bf e}_z + \nu \nabla^2 {\bf u}  \label{eq:DDmomentum}, \\
\nabla \cdot {\bf {u}} &=& 0,  \\
\frac{\partial \tilde{T}}{\partial t} + {\bf u}\cdot\nabla \tilde{T} + \beta_T w   & = & \kappa_T \nabla^2 \tilde{T} \label{eq:DDheat}, \\
\frac{\partial \tilde{C}}{\partial t} + {\bf u}\cdot\nabla \tilde{C} + \beta_C w   & = & \kappa_C \nabla^2 \tilde{C} \label{eq:DDsalt}, 
\end{eqnarray}
where $\alpha_C = \rho_m^{-1} (\partial \rho/\partial C)_T$, and $\kappa_C$ is the compositional diffusivity. For salt water, the diffusivity ratio $\tau = \kappa_C / \kappa_T = O(0.01)$. 

\citet{stern1960sfa} discovered the so-called fingering instability based on the experiment of \citet{Stommelal1956}. This form of double-diffusive convection can take place when temperature (or, more generally, the most rapidly diffusing scalar) is stably stratified while salt (or, the more slowly diffusing scalar) is unstably stratified. Stern argued that a small vertically displaced parcel of fluid would rapidly equilibrate thermally with its surroundings. This reduces the stabilizing impact of the temperature stratification (as it did for the diffusive shear instabilities) and allows in this case the unstable salinity gradient to drive what is essentially haline convection, albeit on small scales. In a footnote, \citet{stern1960sfa} also mentioned that a related oscillatory instability would exist when the stratifications are reversed, namely when temperature is unstably stratified, while salinity is stably stratified. Indeed, in the absence of the temperature field, a vertically displaced parcel of fluid would merely excite stable internal gravity waves in the stably-stratified salinity field. However, in a destabilizing temperature gradient, the diffusive thermal adjustment of the parcel with the ambient temperature provides an additional source of buoyancy that serves to gradually amplify the oscillation (see, e.g. Figure 1 of \cite{Garaud2018}). The linear instability theory for this oscillatory type of double-diffusive convection was presented by \citet{Walin1964}. Double-diffusive instabilities in their various forms (see more on this later) are thought to be significant sources of diapycnal mixing in the tropical ocean, where fingering takes place, and in the polar oceans, where the oscillatory instability and its subcritical manifestations take place \cite{radko2013double}.  

Double-diffusive instabilities were introduced to the astrophysical community in the early 1960s, with the GFD program playing a central role in disseminating the ideas\footnote{Willem Malkus, who worked on the problem early on, introduced fingering instabilities to his colleague Peter Goldreich at UCLA. Goldreich attended GFD in 1966. Spiegel also introduced fingering instability to his colleague Sylvie Vauclair in the 1970s, who went to on discuss their importance in planet formation \cite{Vauclair04}. Meanwhile, Spiegel and Kato were interested in the oscillatory double-diffusive problem, with Kato and Walin both fellows of the GFD program in 1964.}.
In stars, helium and/or other heavier atomic species play the role of salt, while the ionized hydrogen plasma plays the role of the ambient fluid. The fingering instability takes place in stably stratified radiative zones in the presence of an unstable compositional gradient, that could be caused either by material falling onto the surface of the star from accreting planets or from a more evolved binary companion, or created in situ by some nuclear reactions. The instability was, to my knowledge, first invoked as a possible mixing mechanism in stars by \citet{Ulrich1972}, although reference to it in the astrophysical literature dates back to the work of \citet{GoldreichSchubert1967}, who discovered a double-diffusive version of centrifugal instabilities and noted its strong analogy with the thermohaline problem.  
The oscillatory instability on the other hand takes place in regions of the star close to a convective core, where nuclear fusion reactions can create a stabilizing compositional gradient. It was first discussed in the stellar context by \citet{Kato1966} and \citet{Spiegel1969}, who realized its connection with a related problem in stellar astrophysics called semiconvection \cite{SchwarzschildHarm1958}.

I have recently reviewed the topic of double-diffusive instabilities at low Prandtl number in two different venues (the reader is referred to \cite{Garaud2018} for a review addressed to fluid dynamicists, and to \cite{Garaud2020} for a review addressed to stellar astrophysicists). Rather than repeating what can be found elsewhere, I shall therefore focus here on describing the main differences between high and low Prandtl number double-diffusive systems, while garnering insight from what we have just learned about low P\'eclet number flows. 

\subsection{Fingering instabilities} 
\label{sec:fingering}

Using a linear stability analysis, \citet{stern1960sfa} established that the criterion for linear instability to fingering convection depends on the so-called density ratio
\begin{equation}
R_0 = \frac{\alpha_T \beta_T }{\alpha_C \beta_C} .
\end{equation}
A fluid is fingering-unstable provided 
\begin{equation}
1 < R_0 < \frac{\kappa_T}{\kappa_C} = \tau^{-1}.
\end{equation}
Note that the fluid is unstable to multi-component convection when $R_0 \le 1$, and linearly stable if $R_0 >  \tau^{-1}$. The linear stability bound is sharp, and coincides with the energy stability bound \citep{Joseph1976,Balmforthal2006}. In stars, the diffusivity ratio is usually very small, and notably smaller than the kinematic viscosity, so $\tau < Pr \ll 1$. Fingering convection can therefore take place over a huge range of density ratios \cite{schmitt1983}. 
  
Regardless of the Prandtl number, the typical lengthscale associated with the fingering instability is usually of the order of 
\begin{equation}
d = \left( \frac{\kappa_T \nu}{\alpha_T |\beta_T| g }\right)^{1/4} , 
\end{equation}
which is the scale on which the thermal Rayleigh number would be equal to one. In the ocean, the fingers initially develop as thin columns of fluid (hence their name), with alternating warm/salty water flowing down, and cold/fresh water flowing up. The growth rate of fingering modes $\lambda_{fing}$ is the solution of a cubic equation (see, e.g. Radko 2013, equation 2.2), and must usually be computed numerically. The velocity field within the fingers grows exponentially until a secondary shearing instability develops and disrupts them, causing saturation. As discussed by \citet{RadkoSmith2012}  in the geophysical context, and \citet{Brownal2013} in the astrophysical context (see also \cite{denissenkov2010}), it is possible to predict the vertical velocity within the fingers at saturation, $w_{fing}$, simply by requiring a balance between the primary fingering growth rate, and the secondary (parasitic) shear instability growth rate. 

Indeed, prior to the saturation of the primary instability, the flow field associated with the developing fingers is purely vertical, and varies sinusoidally in the horizontal direction on the characteristic lengthscale $d$. The growth rate of the shear instability between the fingers $\lambda_{shear}$ depends on the amplitude of the vertical flow speed $w_{fing}$, and can be computed by linear stability analysis \cite{RadkoSmith2012}. Setting $\lambda_{fing} = K \lambda_{shear}(w_{fing})$ then provides an implicit equation for the finger velocity at saturation, $w_{fing}$. The constant $K$ is finally estimated by fitting the model predictions to the data. The Radko \& Smith model was very successful in predicting the fingering fluxes measured in DNS for a wide range of density ratios, Prandtl number, and diffusivity ratios. 

For geophysically-relevant values of $Pr$, and density ratio of order unity, both DNS and laboratory experiments agree that the turbulent temperature flux can be up to two orders of magnitude larger than the diffusive flux, while the salinity flux can be up to four orders of magnitude larger than the diffusive flux. In the light of what we saw earlier, this implies that the thermal P\'eclet number is large, and that the instability can, in principle, drive the formation of layers and interfaces. This is in fact exactly what happens: for low density ratios, fingering convection is now known to drive the formation of thermohaline staircases, which are stacks of well-mixed fully convective layers separated by thin fingering interfaces. Observed in the ocean \cite{tait1968sot,tait1971ts,schmitt1987c}, and in laboratory experiments \cite{krishnamurti2003double,krishnamurti2009heat} the process by which these layers form was finally clarified by \citet{radko2003mechanism} (see also \cite{Stellmach2011}), who discovered a new mean-field instability he called the $\gamma-$instability, driven by an imbalance between the turbulent salt and temperature fluxes (see \cite{radko2013double} for a review of all processes that lead to the formation of layers in oceanographic double-diffusive convection). These thermohaline staircases are particularly important for diapycnal mixing in the tropical ocean,  where they are observed to significantly increase vertical transport compared with regions where fingering convection alone takes place \citep{schmitt2005enhanced}. 

As $R_0$ increases, the turbulent fluxes of temperature and salinity decrease, and eventually drop to zero at $R_0 = 1/\tau$. For salt water, the turbulence is already very weak (and the thermal P\'eclet number is small) beyond $R_0 \simeq 20$. Based on what we learned in Section \ref{sec:LPNapprox} one may then naturally propose a reduced model for fingering convection using the LPN approximation \citep{Lignieres1999}. Replacing the temperature equation with its approximate form (\ref{eq:LPNbalancedim}), we then have
 \begin{eqnarray}
\frac{\partial{\bf u}}{\partial t} + {\bf u}\cdot\nabla {\bf u} & = & - \frac{1}{\rho_m} \nabla \tilde{p} + \left( \frac{\bar N^2 }{\kappa_T} \nabla^{-2} w - \alpha_C g \tilde{C} \right)  {\bf e}_z + \nu \nabla^2 {\bf u}  \label{eq:LPNDDmomentum}, \\
\nabla \cdot {\bf {u}} &=& 0 ,  \\
\frac{\partial \tilde{C}}{\partial t} + {\bf u}\cdot\nabla \tilde{C} + \beta_C w   & = & \kappa_C \nabla^2 \tilde{C} \label{eq:LPNDDsalt}. 
\end{eqnarray}
These reduced equations were in fact formally derived by \citet{Radko2010} and \citet{xie2017reduced}, using asymptotic expansions of the governing equations for fingering convection near marginal stability (i,e. as $R_0 \rightarrow \tau^{-1}$). However, DNS demonstrate that the region of validity of these equations can be much larger than what the asymptotic theory suggests (e.g. for salt water it is valid for $R_0$ as low as $\sim 20$, rather than in the strict limit $R_0 \rightarrow 100$). 

Meanwhile, in the low Prandtl number limit of stars, \citet{Brownal2013} noted that several simplifications to the \citet{RadkoSmith2012} model can be made. First, it is possible to solve the cubic for the fingering growth rate analytically using an asymptotic expansion for $Pr,\tau \ll 1$, leading to the estimate that 
\begin{equation}
 \lambda_{fing} \simeq \sqrt{\frac{Pr}{R_0} } \frac{\kappa_T}{d^2} +  O(Pr)
\end{equation}
in most of the unstable range except near marginal stability, where $\lambda_{fing}$ drops to zero. 
Secondly, since viscosity is negligible on the fingering scale (again because $Pr \ll 1$), the shear instability growth rate $\lambda_{shear}$ is independent of $\nu$ and can be shown purely on dimensional grounds to be proportional to $ w_{fing} d^{-1}$. Equating the primary and parasitic instability growth rates as before, we find that 
\begin{equation}
w_{fing} \propto d \lambda_{fing} \simeq \sqrt{\frac{Pr}{R_0} } \frac{\kappa_T}{d} \mbox{ when } Pr \ll 1. 
\label{eq:wfing}
\end{equation}
From that, we can estimate a P\'eclet number based on the basic finger properties, to be
\begin{equation}
Pe_{fing} = \frac{d w_{fing}}{\kappa_T}  \simeq \sqrt{\frac{Pr}{R_0} }. 
\label{eq:Pefing}
\end{equation}
Note that a much more detailed asymptotic analysis can be found in \citet{Brownal2013} if required. Since $Pr$ is small and $R_0$ is always larger than one (and can be very large indeed), we see that fingering convection at low Prandtl number always has $Pe_{fing} \ll 1$ at all values of the density ratio. This implies, in particular, that (\ref{eq:LPNDDmomentum})-(\ref{eq:LPNDDsalt}) are {\it always} a good description of fingering convection in stars. It also implies that thermocompositional staircases cannot spontaneously form from the basic fingering instability in that case, since low P\'eclet number flows cannot modify the background temperature profile significantly. This conclusion was already reached by \citet{Traxler2011b}, and further quantified by \citet{Garaudal15b}, by performing a detailed analysis of Radko's mean-field theory \citep{radko2003mechanism} applied to low Prandtl number fingering convection. However, as demonstrated above, we can arrive at the same inevitable conclusion using very simple dimensional arguments.   

The fact that thermocompositional staircases cannot naturally arise in low Prandtl number fingering convection has important implications for stellar structure and evolution. On the one hand, this means that the theory of \citet{Brownal2013}  is sufficient to estimate turbulent mixing by fingering convection in stars. This implies that the turbulent mixing coefficient, for low to moderate values of $R_0$, is 
\begin{equation}
D_{fing} \propto w_{fing} d  \simeq C_{fing} \sqrt{\frac{Pr}{R_0} } \kappa_T \mbox{ for } R_0 \ll \tau^{-1},
\label{eq:Dfing}
\end{equation}
where the constant $C_{fing}$ was estimated by \citet{Brownal2013} to be approximately equal to 10 \cite[see also][]{Garaud2020}.  This formula recovers the functional form of the original model proposed by \citet{Ulrich1972} and is actually quite close in both form and magnitude to the model of  \citet{kippenhahn80}. The mixing coefficient obtained, just as in the case of shear instabilities, is an appropriate estimate for both the turbulent viscosity and the turbulent diffusivity of the scalar $C$ \citep{Garaudal2019} (except in the limit $R_0 \rightarrow \tau^{-1}$, where equation (\ref{eq:Dfing}) is not valid). For values of $R_0$ and $Pr$ appropriate for stellar interiors, $D_{fing}$ ranges from one to three orders of magnitude larger than the microscopic counterparts $\nu$ and $\kappa_C$ (which are of the same order of magnitude, roughly). This has implications, for instance, for observations of the abundance of various chemical species at the surface of Red Giant Branch stars, of planet-bearing stars and of some metal-rich White Dwarf stars (see the review by \cite{Garaud2020} and references therein for more detail).

\subsection{The oscillatory regime} 

\citet{Walin1964} established that the criterion for linear instability to the so-called oscillatory double-diffusive convection (ODDC hereafter; this term was apparently coined by Spiegel) is 
\begin{equation}
1 < R_0^{-1} < \frac{\nu+\kappa_T}{\nu+\kappa_C} = \frac{Pr + 1}{Pr + \tau}.
\label{eq:ODDCcrit}
\end{equation}
The quantity $R_0^{-1}$ is referred to as the "inverse density ratio".  Here the fluid is unstable to multi-component convection when $R_0^{-1} \le 1$, and linearly stable if $R_0^{-1} > (\nu+\kappa_T)/
(\nu+\kappa_C)$. Note how, in contrast with the fingering case, viscosity now affects the linear stability threshold.

For salt water, where $Pr \sim O(10)$, that threshold is very close to one, so the range of inverse density ratios for linear instability is almost negligible. However, as shown by \citet{veronis1965finite} (see also \cite{Proctor1981}), there exists a subcritical branch of instability that persists for $R_0^{-1} \gg (\nu+\kappa_T)/
(\nu+\kappa_C)$. Physically speaking, this is easy to understand: any finite amplitude perturbation that locally reduces the stable salt stratification can allow thermal convection to develop more easily\footnote{Note that several other linear instability mechanisms can also interact with ODDC to give rise to layering, such as the one identified by \citet{Radko2016}.}. Ultimately, the fluid develops one or more convective layers, separated by stably stratified diffusive interfaces. This is the more common form taken by this instability in geophysical flows, as demonstrated in laboratory experiments by \citet{TURNER1965,LindenShirtcliffe1978} and numerical experiments by \citet{carpenter2012simulations}. Thermohaline staircases associated with a stable salt stratification and an unstable temperature stratification are also well documented in the polar oceans and in volcanic lakes \citep{Timmermans2008,Wuest2012}. These staircases typically have an underlying inverse density ratio ranging from 2 to 10, which is stable according to (\ref{eq:ODDCcrit}).

At low Prandtl number, viscosity is negligible on the scales over which thermal diffusion is effective, which means that the marginal stability threshold for the linear oscillatory double-diffusive instability is very large ($R_0^{-1} \sim  O(\tau^{-1}) \sim10^6$ or larger, as in the fingering case). As such, this instability is dynamically relevant, by contrast with the geophysical context where it is not. Almost any region that has an unstable potential temperature gradient can therefore be the seat of ODDC, even when the stabilizing compositional stratification is extremely strong \citep{Kato1966,Spiegel1969}. 

The nonlinear saturation of the primary instability in ODDC remains an open question. Ad-hoc models for the turbulent compositional flux induced by ODDC were proposed by \citet{Stevenson1977} and \citet{Langer1983}, but neither appear to fit recent DNS results of \citet{Mirouh2012} (see \cite{Garaud2020} for a comparison). For low inverse density ratios, my former graduate student Ryan Moll demonstrated in his MS Thesis that a model similar to the \citet{Brownal2013} model for fingering convection (i.e. equating the primary instability growth rate to the growth rate of parasitic shear instabilities) can explain the turbulent flux data. At larger inverse density ratios, however, this model does not work and largely under-predicts the turbulent fluxes. The instability appears to saturate instead through the generation and interaction with large-scale shearing modes (often called "jets"), as demonstrated by \citet{Paparella2002} (another GFD project) using a truncated modal analysis, and by \citet{Moll2016} using 3D DNS. It is interesting to note that in the same limit thermal diffusion becomes dominant, so the truncated model (\ref{eq:LPNDDmomentum})-(\ref{eq:LPNDDsalt}) is expected to be a good approximation for the dynamics of the system. To my knowledge, this has not been explored yet. In short, a comprehensive theory of the nonlinear saturation of ODDC is still lacking, but several encouraging avenues exist that ought to be further investigated.  

An important outcome of the numerical experiments performed by my research group over the years, starting with the exploratory work of \citet{rosenblumal2011}, is the discovery that ODDC at low $R_0^{-1}$ undergoes a spontaneous transition to layered convection through Radko's $\gamma$-instability (see above). The $\gamma$-instability takes place whenever the ratio $\gamma$ of the total buoyancy flux due to compositional transport, to the total buoyancy flux due to temperature transport, is a decreasing function of $R_0^{-1}$. As demonstrated by \citet{Mirouh2012}, the range of inverse density ratios for which this is the case increases as the Prandtl number and diffusivity ratio both decrease, suggesting that layered double-diffusive convection should be prevalent in stellar interiors (see the reviews \cite{Garaud2018,Garaud2020}), especially in the vicinity of convective cores \citep{MooreGaraud2016}.

\begin{figure}[h]
\includegraphics[width=\textwidth]{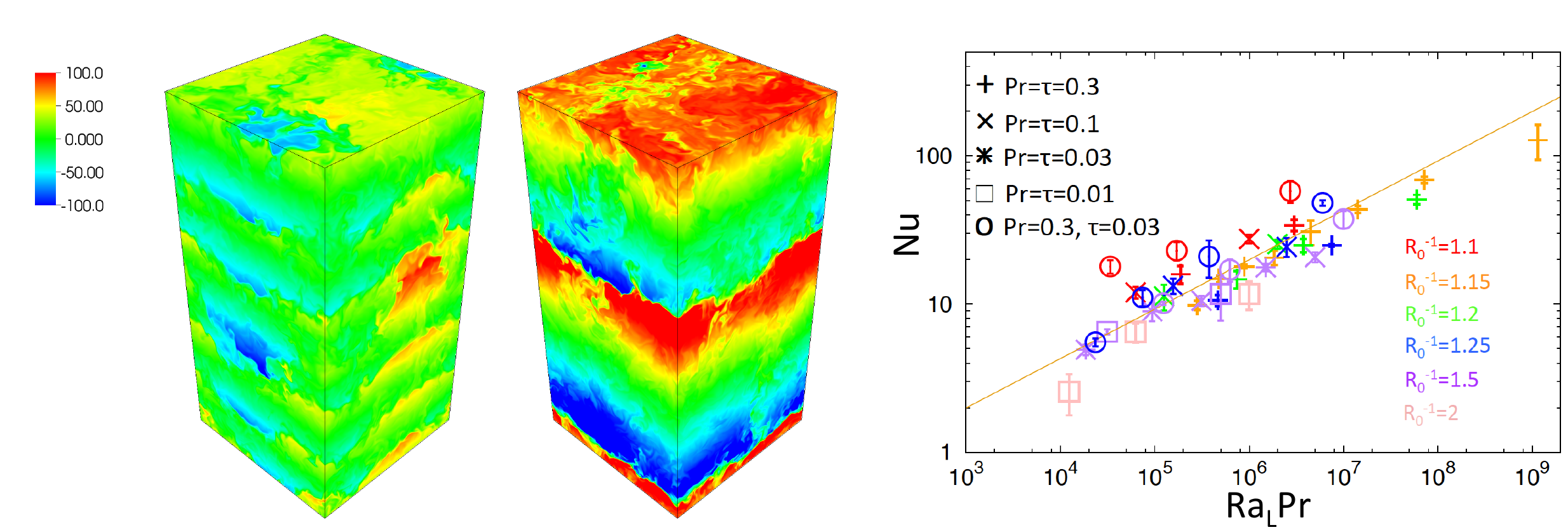}
\caption{Left: Snapshots of the compositional perturbations $\tilde C$ in a DNS of ODDC at $R_0^{-1} = 1.15$, $Pr = \tau = 0.3$, presented in \citet{Woodal13}, at two instants in time, showing the formation of layers and their subsequent mergers. Right: Nusselt number as a function of $Ra_L Pr$ in many different DNS of ODDC at varying $R_0^{-1}$, $Pr$ and $\tau$, showing the scaling $Nu \propto (Ra_L Pr)^{1/3}$. Figure adapted from \citet{Garaud2018}.}
\label{fig:layered}
\end{figure}

\citet{Woodal13} and \citet{Mollal2017} performed a series of DNS of ODDC in the layered regime, at low Prandtl number. We found that in a staircase composed of multiple layers with roughly equal heights $L$, subject to a mean potential temperature stratification $\beta_T$ and a mean compositional stratification $\beta_C$ (so the potential temperature and compositional jumps across each interface are $\Delta T = |\beta_T L|$ and $\Delta C = |\beta_C L|$, respectively), the Nusselt number is proportional to $(Ra_L Pr)^{1/3}$, where $Ra_L$ is the layer-based thermal Rayleigh number
\begin{equation}
Ra_L = \frac{\alpha_T |\beta_T| g L^4}{\kappa_T \nu} .
\end{equation}
The heat flux might also depend more weakly on the inverse density ratio and the diffusivity ratio, but the available data is too limited to conclusively propose any scaling (see figure \ref{fig:layered}). 
This scaling law is consistent with the notion that, in the absence of solid boundaries (and their associated viscous boundary layers), the heat flux should become independent of $Pr$ for asymptotically low $Pr$.

The simulations of \citet{Woodal13} and \citet{Mollal2017} are still preliminary, but are the only ones to my knowledge to study layered double-diffusive convection at low Prandtl number in the absence of solid boundaries (which do not exist in stars). Simulations in a bounded domain (between solid plates) were presented by \citet{Biello2001}, \citet{ZaussingerSpruit2013} and \citet{ZaussingerKupka2019}, and behave quite differently, which is expected. Generally speaking, much more remains to be done to understand and fully characterize layered double-diffusive convection a low Prandtl number. Taller and wider computational domains, as well as significantly higher resolution, will be needed to probe a larger region of parameter space in terms of Pr, $\tau$, $R_0^{-1}$ and layer height, to confirm or challenge the scaling laws proposed by \citet{Woodal13}. It also would be particularly interesting to see whether one eventually recovers the Ultimate Regime of convection as $Ra_L$ increases. Finally, note that the simulations of \citet{Woodal13} at low $R_0^{-1}$ suggest that thermocompositional staircases are not a stable configuration in the long term, because individual layers have a tendency to merge over time until a single fully convective layer remains (see figure \ref{fig:layered}). Whether this is an artifact of the boundary conditions used, or a genuine property of layered double-diffusive convection, needs to be established. 

The answer to these questions have important implications for stellar and planetary astrophysics. Layered double-diffusive convection in the vicinity of the convective cores of intermediate and high-mass stars, as demonstrated by \citet{MooreGaraud2016}, helps transport hydrogen into the core, which, as mentioned in Section \ref{sec:intro}, fuels its nuclear reactions, prolongs the lifetime of the star, and increases the size of the core prior to its end-of-life stage (supernova or red giant). This in turn impacts the ultimate redistribution of metal-enriched material in the host galaxy, with implications for star formation and cosmology. In the interior of giant planets, \citet{Mollal2017} (see also \cite{StevensonSalpeter1977,Guillot2004}) showed that the presence or absence of double-diffusive layers can control the rate at which the convective envelope erodes the primordial rocky or water-rich core. Again, this has potentially observable consequences, and needs to be taken into account in models of the formation and evolution of planets. More examples of the importance of ODDC in astrophysics are discussed in \citet{Garaud2020}. 

\section{Magnetic fields and rotation}
\label{sec:magrot}

Before concluding, a few remarks are perhaps in order. It was my goal in this review to present a few instabilities that are of particular interest to both geophysical fluid dynamicists and to the stellar astrophysics community, while emphasizing fundamental differences in the emergent turbulence that are due solely to the fluid's Prandtl number. In choosing to focus on a few selected topics only, I have had to ignore many others that also play an important role in both stars and in the Earth's oceans, atmosphere, and/or molten interior. These include, in no specific order, topics such as centrifugal instabilities, large-scale meridional circulations, gravity waves and Rossby waves, penetrative convection, and the generation of magnetic fields by dynamo action (among others). More importantly, I have neglected to include the effects of rotation and magnetic fields on the three types of instabilities discussed in this review. This choice was made for pedagogical purposes, but by doing so I have vastly oversimplified the physics to the point that many of the results presented cannot be directly applied to model stellar interiors. In what follows, I provide a very brief glimpse into the various ways in which rotation and magnetic fields can change the results presented in the previous sections. 

\subsection{Rotation}

All astrophysical and geophysical systems are rotating to a greater or lesser degree, and rotation needs to be taken into account whenever the Rossby number of the flow, defined as $Ro = U/L\Omega$ (where $U$ and $L$ here are the characteristic velocity and lengthscale of the dominant eddies, respectively, and $\Omega$ is the local rotation rate), is of order unity or lower. Conservation of angular momentum in an inviscid rotating system strongly constrains the range of dynamics allowed. On the one hand, gradients of angular momentum can have a stabilizing or destabilizing effect on certain types of perturbations, and therefore constrain both the linear and nonlinear development of instabilities. On the other hand, the Taylor-Proudman constraint forces all components of a very low Rossby number flow to be invariant along the rotation axis, so the resulting turbulent dynamics become almost two-dimensional. The effect of rotation on all three kinds of instabilities discussed in this review is generally relatively well understood at $Pr \sim O(1)$, but, to my knowledge, there are very few instances in which the regime combining $Ro < 1$ and $Pr \ll 1$ has been considered.  

Significant progress has recently been made in improving our understanding of the impact of rotation on Rayleigh-B\'enard convection at $Pr \sim O(1)$, thanks to a combination of numerical experiments and laboratory experiments, as well as the development of reduced asymptotic models (see reviews of the topic in e.g. \cite{Chengal2018,PlumleyJulien2019}). A key result of the last decade is the identification of a new rotationally-constrained, yet fully-turbulent convective regime, that emerges past the onset of convective instability. In this regime, the Nusselt number scales as $Nu \propto Pr^\gamma (RaE^{4/3})^\alpha$, where $E =  \nu /2 \Omega H^2$ is the Ekman number, and $\alpha$ and $\gamma$ are two exponents that depend on the nature of the system boundaries (no slip vs. stress free). Whether this regime exhibits similar scaling laws at low Prandtl number remains to be established, and it will be interesting to see whether some of Spiegel's \cite{Spiegel1962} asymptotic arguments apply for rotating convection.

Relatively little is known about the effects of rotation on double-diffusive instabilities at low Prandtl number (although there are some preliminary studies  \cite{SenguptaGaraud2018,MollGaraud2017}), with one notable exception, which is the Goldreich-Schubert-Fricke (GSF) instability \cite{GoldreichSchubert1967,Fricke1968}. The GSF instability is a doubly-diffusive centrifugal instability (where the angular-momentum gradient is destabilizing, and the thermal stratification is stabilizing) that is increasingly recognized as an important source of angular momentum transport in stars. It bears many similarities with fingering convection, and has recently been studied in depth by Barker, Jones and Tobias \cite{Barkeral2019,Barkeral2020}. By nature, the GSF instability exists {\it only} when the Prandtl number is small, and, in two dimensions, is an almost exact analog of the fingering instability. As such, it lends itself well to some of the asymptotic arguments and reduced modeling described in Section \ref{sec:fingering}. 

Finally, to my knowledge there has not yet been any systematic analysis of the influence of rotation on the nonlinear development of stratified shear instabilities (vertical or horizontal) at low Prandtl number. Linear stability analyses demonstrate the existence of several modes of instability (including, depending on the model setup, baroclinic modes, GSF modes, and shearing modes that are rotationally constrained, see for instance \cite{Rashidal2008,Parkal2020}). The recent MS thesis of my former student Eonho Chang, which is currently under preparation for publication, presents a preliminary analysis of the nonlinear development of low P\'eclet number vertical shear instabilities in the presence of rotation, and their interaction with centrifugal instabilities. His work reveals the existence of several different parameter regimes, depending on the relative strengths of the rotation and the stratification, including some that are, respectively, rotationally-dominated, shear-dominated, or controlled by the GSF instability, and some that exhibit quasi-periodic excursions from one regime to the other. Since many of these regimes are likely relevant to stellar evolution, we will need to dedicate time and resources in the future to better characterize and quantify their properties. 

\subsection{Magnetic fields}

Magnetic fields are not thought to be relevant in the dynamics of the Earth's oceans and atmosphere, but are fundamentally important in stellar interiors.  Indeed, the very high conductivity of the plasma usually implies that all but the smallest-scale or weakest fluid motions have a large magnetic Reynolds number, which in turn implies that the flow can exponentially amplify magnetic fields by dynamo action (see, e.g. \cite{Moffatt1978} and the excellent recent lecture by Fran\c{c}ois Rincon \cite{Rincon2019}). The energy of the magnetic field usually grows to reach some fraction of the total turbulent kinetic energy of the flow and the Lorentz force becomes dynamically significant, modifying the properties of the turbulence in a way that saturates the dynamo instability. We therefore see that, by nature, magnetic fields necessarily play a leading-order role in turbulent stellar plasmas, and should never be ignored: if a dynamo is excited, then the transport properties of the turbulence (momentum transport, heat transport, compositional transport) are likely affected by the field.

Convective dynamos are by far the most widely-discussed types of stellar dynamos, and are usually the source of a star's observable magnetic field \cite{Mestel1999}. Since convection draws its energy from the unstable stratification (a finite reservoir of potential energy), and since the field is amplified by converting kinetic energy into magnetic energy, the dynamo process might naturally be expected to reduce the efficiency of convective heat transport somewhat. However, \citet{Yanal2021} recently showed that the effect is fairly small, and that the presence of a small-scale dynamo does not seem to affect the $Nu(Ra)$ scaling law of hydrodynamic convection. This likely explains why the standard mixing length theory of B\"ohm-Vitense (see Section \ref{sec:lowPrapprox}) is generally quite successful at modeling stellar convection, despite ignoring magnetic fields entirely. It is worth noting, however, that saturation of the convective instability in Spiegel's asymptotically low Prandtl number regime \cite{Spiegel1962} is due to the turbulent stresses in the momentum equation (rather than convective fluxes in the temperature equation, which is the more classic scenario). As such, the dynamo field may influence the heat flux much more significantly in the limited region of parameter space where his equations are valid, namely $Ra_c  < Ra \ll Pr^{-1}$, see Section \ref{sec:lowPrapprox}. 

For instabilities taking place in stellar radiative zones (e.g. stratified shear instabilities and double-diffusive instabilities), one needs to distinguish between the impact of the small-scale dynamo field generated locally by the turbulence resulting from the instability itself, and the impact of a large-scale "external" magnetic field that exists independently of the instability. These two cases are quite different, as the former only affects the nonlinear development of the instability, while the latter can also impact its initial exponential growth. 

Relatively little is known so far about the impact of magnetic fields on double-diffusive instabilities in stars. From the perspective of linear theory, a large-scale externally imposed magnetic field quenches perturbations that vary along the direction of the field, but leaves those that are invariant along the field free to grow   normally \cite{CharbonnelZahn2007b}. As such, the fastest-growing modes in both fingering and ODDC instabilities are unaffected. Beyond linear theory, however, the story becomes a lot more complex. To my knowledge, there is no published study on the impact of magnetic fields on ODDC, so the question remains entirely open. In the fingering case, on the other hand, recent work by Peter Harrington and I \cite{HarringtonGaraud2019} has demonstrated that the presence of a large-scale vertical magnetic field (i.e. aligned with gravity) can actually {\it enhance} vertical compositional transport by fingering convection, because it suppresses the parasitic instabilities that normally saturate the fingers. However, it is not clear what would  happen if the external field were inclined, or in the absence of an externally imposed large-scale field. The question of momentum transport by magnetized fingering convection also remains to be addressed.  

Finally, studying (or even merely reviewing) the interaction between an externally imposed magnetic field and stratified shear instabilities is a momentous task, 
owing to the large dimensionality of parameter space. As mentioned in Section \ref{sec:intro}, instabilities arising from the combination of magnetic fields and shear can take many different forms depending on the model geometry. A perhaps more approachable question would be to quantify the impact of a locally-generated dynamo field on the transport properties of the shear-induced turbulence presented in Section \ref{sec:stratshear}, but to my knowledge, even that still remains an outstanding question. Small-scale turbulence seems to be able to drive a small-scale dynamo in stellar radiative zones provided the turbulence is sufficiently strong (i.e. the magnetic Reynolds number $Rm = UL/\eta$ is sufficiently large, where $\eta$ is the magnetic diffusivity) and sufficiently three-dimensional \cite{Skoutneval2021}. This dynamo field can therefore quite plausibly affect the predicted turbulent mixing coefficients discussed in Section \ref{sec:stratshear}. As a matter of fact, there is a growing amount of evidence suggesting that the Maxwell stresses associated with the small-scale dynamo sometimes have a tendency to {\it oppose} the Reynolds stresses of the turbulence, even when the field saturates substantially below equipartition (see, e.g. \cite{TDH2007} for evidence in 2D $\beta$-plane simulations with a weak mean field and \cite{Gilman1983,Varela2016} in 3D simulations of magnetized rotating convection in a spherical shell). As such, the results presented in Section \ref{sec:stratshear} on {\it momentum} transport in low Prandtl number stratified turbulence are most likely not applicable in magnetized plasmas.

\section{Perspective: the importance of multidisciplinary programs} 
\label{sec:ccl}

 As we discovered, an asymptotically small Prandtl number causes various instabilities that are common in stellar interiors to behave very differently from their geophysical, moderate-to-high Prandtl number counterparts. As such, the field of stellar fluid dynamics is a goldmine of interesting projects for young scientists, especially thanks to high-performance computing. And yet, it remains the case today that most stellar astrophysicists do not receive a rigorous training in fluid dynamics, and so these enticing projects are just waiting to be investigated by those who do. 
 
In this respect, I cannot overstate the fundamental role that multidisciplinary science programs, such as the Woods Hole Geophysical Fluid Dynamics summer program (GFD), have played in moving the field of stellar astrophysics forward. 
The modern view of stellar fluid dynamics presented in this paper began with a meeting between Ed Spiegel and Willem Malkus, which eventually led to Spiegel's participation as one of the seven founding members of GFD. The program founders believed in the importance of multidisciplinary research, around the central themes of applied mathematics and fluid dynamics, and GFD was never just about geophysical flows. They continued to invite, year after year, many of the foremost astrophysical fluid dynamicists, as well as aspiring graduate students and postdocs in astrophysics, to join them for a summer or more. The early years welcomed giants of the field, such as my advisors Douglas Gough and Nigel Weiss, as well as Steven Balbus, Peter Goldreich, Andy Ingersoll, Bob Stein, Jean-Paul Zahn, and many others. They went on to use these more rigorous fluid dynamical approaches in their own research on stellar and planetary fluid dynamics, and in turn inspired their postdocs and students (including myself) to do the same.

Despite this, stellar fluid dynamics has remained a fairly marginal aspect of stellar astrophysics, not least because fluid motions in stars (other than the Sun) are quite difficult to observe. As such, the validity of a turbulent mixing prescription, or of a model for wave-induced transport or for large-scale flows, can only be tested indirectly by studying their impact on the few observable surface properties of the star (e.g. chemical species abundances or surface rotation rate, among others). Since these properties depend on a combination of many different processes, some known and others most likely unknown, it is rarely possible to disentangle their contributions. Thankfully, asterosexismology, combined with precision astrometry and data science, are slowly beginning to change this status quo, and stellar astrophysics will need to adapt accordingly by finally accepting that stars are fluid objects that are at least as dynamically complex as our own Earth's oceans and atmosphere.  

\acknowledgements

P.G. thanks the National Science Foundation for funding the research described in this review, including grants number 0807672, 0933759, 1211394, 1412951, 1517927, 1814327 and 1908338. She also thanks the Office of Naval Research and the National Science Foundation for supporting the Woods Hole GFD program. 

\bibliographystyle{unsrtnat}
\bibliography{NSF-bib}

\providecommand{\noopsort}[1]{}\providecommand{\singleletter}[1]{#1}%
\begin{thebibliography}{144}
\providecommand{\natexlab}[1]{#1}
\providecommand{\url}[1]{\texttt{#1}}
\expandafter\ifx\csname urlstyle\endcsname\relax
  \providecommand{\doi}[1]{doi: #1}\else
  \providecommand{\doi}{doi: \begingroup \urlstyle{rm}\Url}\fi

\bibitem[{Cox} and {Giuli}(1968)]{CoxGiuli1968}
J.~P. {Cox} and R.~T. {Giuli}.
\newblock \emph{{Principles of stellar structure}}.
\newblock 1968.

\bibitem[{Kippenhahn} et~al.(2012){Kippenhahn}, {Weigert}, and
  {Weiss}]{KippenhahnWeigert2012}
Rudolf {Kippenhahn}, Alfred {Weigert}, and Achim {Weiss}.
\newblock \emph{{Stellar Structure and Evolution}}.
\newblock 2012.
\newblock \doi{10.1007/978-3-642-30304-3}.

\bibitem[{Stix}(2004)]{Stix2004}
Michael {Stix}.
\newblock \emph{{The sun : an introduction}}.
\newblock 2004.

\bibitem[{Kato} and {Fukue}(2020)]{KatoFukue2020}
Shoji {Kato} and Jun {Fukue}.
\newblock \emph{{Fundamentals of Astrophysical Fluid Dynamics; Hydrodynamics,
  Magnetohydrodynamics, and Radiation Hydrodynamics}}.
\newblock 2020.
\newblock \doi{10.1007/978-981-15-4174-2}.

\bibitem[{Garaud}(2018)]{Garaud2018}
Pascale {Garaud}.
\newblock {Double-Diffusive Convection at Low Prandtl Number}.
\newblock \emph{Annual Review of Fluid Mechanics}, 50\penalty0 (1):\penalty0
  275--298, January 2018.
\newblock \doi{10.1146/annurev-fluid-122316-045234}.

\bibitem[{Garaud}(2020{\natexlab{a}})]{Garaud2020}
Pascale {Garaud}.
\newblock {Double-diffusive processes in stellar astrophysics}.
\newblock In Michel {Rieutord}, Isabelle {Baraffe}, and Yveline {Lebreton},
  editors, \emph{Multi-Dimensional Processes In Stellar Physics}, page~13,
  January 2020{\natexlab{a}}.

\bibitem[{Zahn}(1974)]{Zahn1974}
J.-P. {Zahn}.
\newblock {Rotational instabilities and stellar evolution}.
\newblock In P.~{Ledoux}, A.~{Noels}, and A.~W. {Rodgers}, editors,
  \emph{Stellar Instability and Evolution}, volume~59 of \emph{IAU Symposium},
  pages 185--194, 1974.

\bibitem[{Spruit} and {Knobloch}(1984)]{SpruitKnobloch1984}
H.~C. {Spruit} and E.~{Knobloch}.
\newblock {Baroclinic instability in stars}.
\newblock \emph{A\&A}, 132\penalty0 (1):\penalty0 89--96, March 1984.

\bibitem[{Mestel}(1999)]{Mestel1999}
Leon {Mestel}.
\newblock \emph{{Stellar magnetism}}.
\newblock 1999.

\bibitem[{Tayler}(1973)]{Tayler11973}
R.~J. {Tayler}.
\newblock {The adiabatic stability of stars containing magnetic
  fields-I.Toroidal fields}.
\newblock \emph{MNRAS}, 161:\penalty0 365, January 1973.
\newblock \doi{10.1093/mnras/161.4.365}.

\bibitem[{Markey} and {Tayler}(1973)]{Tayler21973}
P.~{Markey} and R.~J. {Tayler}.
\newblock {The adiabatic stability of stars containing magnetic fields. II.
  Poloidal fields}.
\newblock \emph{MNRAS}, 163:\penalty0 77--91, March 1973.
\newblock \doi{10.1093/mnras/163.1.77}.

\bibitem[{Dikpati} and {Gilman}(1999)]{DikpatiGilman1999}
Mausumi {Dikpati} and Peter~A. {Gilman}.
\newblock {Joint Instability of Latitudinal Differential Rotation and
  Concentrated Toroidal Fields below the Solar Convection Zone}.
\newblock \emph{ApJ}, 512\penalty0 (1):\penalty0 417--441, February 1999.
\newblock \doi{10.1086/306748}.

\bibitem[{Cally} et~al.(2003){Cally}, {Dikpati}, and {Gilman}]{Callyal2003}
Paul~S. {Cally}, Mausumi {Dikpati}, and Peter~A. {Gilman}.
\newblock {Clamshell and Tipping Instabilities in a Two-dimensional
  Magnetohydrodynamic Tachocline}.
\newblock \emph{ApJ}, 582\penalty0 (2):\penalty0 1190--1205, January 2003.
\newblock \doi{10.1086/344746}.

\bibitem[{Acheson}(1979)]{Acheson1979}
D.~J. {Acheson}.
\newblock {Instability by magnetic buoyancy.}
\newblock \emph{Solar Physics}, 62\penalty0 (1):\penalty0 23--50, May 1979.
\newblock \doi{10.1007/BF00150129}.

\bibitem[{Schmitt} and {Rosner}(1983)]{SchmittRosner1983}
J.~H.~M.~M. {Schmitt} and R.~{Rosner}.
\newblock {Doubly diffusive magnetic buoyancy instability in the solar
  interior}.
\newblock \emph{APJ}, 265:\penalty0 901--924, February 1983.
\newblock \doi{10.1086/160734}.

\bibitem[{Spiegel} and {Veronis}(1960)]{SpiegelVeronis1960}
E.~A. {Spiegel} and G.~{Veronis}.
\newblock {On the Boussinesq Approximation for a Compressible Fluid.}
\newblock \emph{ApJ}, 131:\penalty0 442, March 1960.

\bibitem[{Vitense}(1953)]{Vitense1953}
E.~{Vitense}.
\newblock {Die Wasserstoffkonvektionszone der Sonne. Mit 11 Textabbildungen}.
\newblock \emph{Zeit. Astrophys.}, 32:\penalty0 135, January 1953.

\bibitem[{Priestley}(1954)]{Priestley1954}
C.~H.~B. {Priestley}.
\newblock {Convection from a Large Horizontal Surface}.
\newblock \emph{Australian Journal of Physics}, 7:\penalty0 176, March 1954.
\newblock \doi{10.1071/PH540176}.

\bibitem[{Malkus}(1954)]{Malkus1954}
W.~V.~R. {Malkus}.
\newblock {The Heat Transport and Spectrum of Thermal Turbulence}.
\newblock \emph{Proceedings of the Royal Society of London Series A},
  225\penalty0 (1161):\penalty0 196--212, August 1954.
\newblock \doi{10.1098/rspa.1954.0197}.

\bibitem[Howard(1966)]{Howard1966}
Louis~N Howard.
\newblock Convection at high rayleigh number.
\newblock In \emph{Applied Mechanics}, pages 1109--1115. Springer, 1966.

\bibitem[Niemela and Sreenivasan(2003)]{niemelasreenivasan2003}
J.~J. Niemela and K.~R. Sreenivasan.
\newblock Confined turbulent convection.
\newblock \emph{JFM}, 481:\penalty0 355?384, 2003.
\newblock \doi{10.1017/S0022112003004087}.

\bibitem[{He} et~al.(2012){He}, {Funfschilling}, {Nobach}, {Bodenschatz}, and
  {Ahlers}]{Heal2012}
Xiaozhou {He}, Denis {Funfschilling}, Holger {Nobach}, Eberhard {Bodenschatz},
  and Guenter {Ahlers}.
\newblock {Transition to the Ultimate State of Turbulent Rayleigh-B{\'e}nard
  Convection}.
\newblock \emph{PRL}, 108\penalty0 (2):\penalty0 024502, January 2012.
\newblock \doi{10.1103/PhysRevLett.108.024502}.

\bibitem[{Doering}(2020)]{Doering2020}
Charles~R. {Doering}.
\newblock {Absence of Evidence for the Ultimate State of Turbulent
  Rayleigh-B{\'e}nard Convection}.
\newblock \emph{PRL}, 124\penalty0 (22):\penalty0 229401, June 2020.
\newblock \doi{10.1103/PhysRevLett.124.229401}.

\bibitem[{Spiegel}(1962)]{Spiegel1962}
E.~A. {Spiegel}.
\newblock {Thermal turbulence at very small Prandtl number}.
\newblock \emph{J. Geophys. Res.}, 67:\penalty0 3063--3070, July 1962.

\bibitem[{Spiegel}(1963)]{Spiegel1963}
Edward~A. {Spiegel}.
\newblock {A Generalization of the Mixing-Length Theory of Turbulent
  Convection.}
\newblock \emph{ApJ}, 138:\penalty0 216, July 1963.
\newblock \doi{10.1086/147628}.

\bibitem[{Kraichnan}(1962)]{Kraichnan1962}
Robert~H. {Kraichnan}.
\newblock {Turbulent Thermal Convection at Arbitrary Prandtl Number}.
\newblock \emph{Physics of Fluids}, 5\penalty0 (11):\penalty0 1374--1389,
  November 1962.
\newblock \doi{10.1063/1.1706533}.

\bibitem[Roche et~al.(2001)Roche, Castaing, Chabaud, and
  H{\'e}bral]{Rocheal2001}
P.-E. Roche, B.~Castaing, B.~Chabaud, and B.~H{\'e}bral.
\newblock Observation of the $\frac{1}{2}$ power law in rayleigh-b{\'e}nard
  convection.
\newblock \emph{Phys. Rev. E}, 63:\penalty0 045303, Mar 2001.
\newblock \doi{10.1103/PhysRevE.63.045303}.
\newblock URL \url{https://link.aps.org/doi/10.1103/PhysRevE.63.045303}.

\bibitem[Lepot et~al.(2018)Lepot, Auma{\^\i}tre, and Gallet]{Lepotal2018}
Simon Lepot, S{\'e}bastien Auma{\^\i}tre, and Basile Gallet.
\newblock Radiative heating achieves the ultimate regime of thermal convection.
\newblock \emph{Proceedings of the National Academy of Sciences}, 115\penalty0
  (36):\penalty0 8937--8941, 2018.
\newblock ISSN 0027-8424.
\newblock \doi{10.1073/pnas.1806823115}.
\newblock URL \url{https://www.pnas.org/content/115/36/8937}.

\bibitem[{Bouillaut} et~al.(2019){Bouillaut}, {Lepot}, {Auma{\^\i}tre}, and
  {Gallet}]{Bouillautal2019}
Vincent {Bouillaut}, Simon {Lepot}, S{\'e}bastien {Auma{\^\i}tre}, and Basile
  {Gallet}.
\newblock {Transition to the ultimate regime in a radiatively driven convection
  experiment}.
\newblock \emph{JFM}, 861:\penalty0 R5, February 2019.
\newblock \doi{10.1017/jfm.2018.972}.

\bibitem[{Miquel} et~al.(2020){Miquel}, {Bouillaut}, {Aumaitre}, and
  {Gallet}]{Miquelal2020}
Benjamin {Miquel}, Vincent {Bouillaut}, Sebastien {Aumaitre}, and Basile
  {Gallet}.
\newblock {On the role of the Prandtl number in convection driven by heat
  sources and sinks}.
\newblock \emph{arXiv e-prints}, art. arXiv:2006.07109, June 2020.

\bibitem[{Thual}(1992)]{Thual1992}
O.~{Thual}.
\newblock {Zero-Prandtl-number convection}.
\newblock \emph{JFM}, 240:\penalty0 229--258, January 1992.
\newblock \doi{10.1017/S0022112092000089}.

\bibitem[{Malkus} and {Veronis}(1958)]{MalkusVeronis1958}
W.~V.~R. {Malkus} and G.~{Veronis}.
\newblock {Finite amplitude cellular convection}.
\newblock \emph{Journal of Fluid Mechanics}, 4:\penalty0 225--260, January
  1958.
\newblock \doi{10.1017/S0022112058000410}.

\bibitem[{Ledoux} et~al.(1961){Ledoux}, {Schwarzschild}, and
  {Spiegel}]{Ledouxal1961}
P.~{Ledoux}, M.~{Schwarzschild}, and E.~A. {Spiegel}.
\newblock {On the Spectrum of Turbulent Convection.}
\newblock \emph{ApJ}, 133:\penalty0 184, January 1961.
\newblock \doi{10.1086/147015}.

\bibitem[{Howard}(1963)]{Howard1963}
L.~N. {Howard}.
\newblock {Heat transport by turbulent convection}.
\newblock \emph{JFM}, 17:\penalty0 405--432, January 1963.
\newblock \doi{10.1017/S0022112063001427}.

\bibitem[Doering and Constantin(1996)]{DoeringConstantin1996}
Charles~R. Doering and Peter Constantin.
\newblock Variational bounds on energy dissipation in incompressible flows.
  iii. convection.
\newblock \emph{Phys. Rev. E}, 53:\penalty0 5957--5981, Jun 1996.
\newblock \doi{10.1103/PhysRevE.53.5957}.
\newblock URL \url{https://link.aps.org/doi/10.1103/PhysRevE.53.5957}.

\bibitem[{Ligni{\`e}res}(1999)]{Lignieres1999}
F.~{Ligni{\`e}res}.
\newblock {The small-P{\'e}clet-number approximation in stellar radiative
  zones}.
\newblock \emph{A.\&A.}, 348:\penalty0 933--939, August 1999.

\bibitem[{Thompson} et~al.(1996){Thompson}, {Toomre}, {Anderson}, {Antia},
  {Berthomieu}, {Burtonclay}, {Chitre}, {Christensen-Dalsgaard}, {Corbard}, {De
  Rosa}, {Genovese}, {Gough}, {Haber}, {Harvey}, {Hill}, {Howe}, {Korzennik},
  {Kosovichev}, {Leibacher}, {Pijpers}, {Provost}, {Rhodes}, {Schou}, {Sekii},
  {Stark}, and {Wilson}]{Thompson-etal96}
M.~J. {Thompson}, J.~{Toomre}, E.~R. {Anderson}, H.~M. {Antia},
  G.~{Berthomieu}, D.~{Burtonclay}, S.~M. {Chitre}, J.~{Christensen-Dalsgaard},
  T.~{Corbard}, M.~{De Rosa}, C.~R. {Genovese}, D.~O. {Gough}, D.~A. {Haber},
  J.~W. {Harvey}, F.~{Hill}, R.~{Howe}, S.~G. {Korzennik}, A.~G. {Kosovichev},
  J.~W. {Leibacher}, F.~P. {Pijpers}, J.~{Provost}, E.~J. {Rhodes}, Jr.,
  J.~{Schou}, T.~{Sekii}, P.~B. {Stark}, and P.~R. {Wilson}.
\newblock {Differential Rotation and Dynamics of the Solar Interior}.
\newblock \emph{Science}, 272:\penalty0 1300--1305, May 1996.
\newblock \doi{10.1126/science.272.5266.1300}.

\bibitem[{Aerts} et~al.(2019){Aerts}, {Mathis}, and {Rogers}]{Aertsal2019}
Conny {Aerts}, St{\'e}phane {Mathis}, and Tamara~M. {Rogers}.
\newblock {Angular Momentum Transport in Stellar Interiors}.
\newblock \emph{Ann. Rev. Astron. Astrophys.}, 57:\penalty0 35--78, August
  2019.
\newblock \doi{10.1146/annurev-astro-091918-104359}.

\bibitem[{Charbonneau} et~al.(1999){Charbonneau}, {Christensen-Dalsgaard},
  {Henning}, {Larsen}, {Schou}, {Thompson}, and {Tomczyk}]{Charbonneaual99}
P.~{Charbonneau}, J.~{Christensen-Dalsgaard}, R.~{Henning}, R.~M. {Larsen},
  J.~{Schou}, M.~J. {Thompson}, and S.~{Tomczyk}.
\newblock {Helioseismic Constraints on the Structure of the Solar Tachocline}.
\newblock \emph{ApJ}, 527:\penalty0 445--460, December 1999.

\bibitem[{Larson} and {Schou}(2018)]{LarsonSchou2018}
Timothy~P. {Larson} and Jesper {Schou}.
\newblock {Global-Mode Analysis of Full-Disk Data from the Michelson Doppler
  Imager and the Helioseismic and Magnetic Imager}.
\newblock \emph{Solar Physics}, 293\penalty0 (2):\penalty0 29, February 2018.
\newblock \doi{10.1007/s11207-017-1201-5}.

\bibitem[{Richardson}(1920)]{Richardson1920}
L.~F. {Richardson}.
\newblock {The Supply of Energy from and to Atmospheric Eddies}.
\newblock \emph{Royal Society of London Proceedings Series A}, 97:\penalty0
  354--373, July 1920.
\newblock \doi{10.1098/rspa.1920.0039}.

\bibitem[{Miles}(1961)]{Miles61}
J.~W. {Miles}.
\newblock {On the stability of heterogeneous shear flows}.
\newblock \emph{JFM}, 10:\penalty0 496--508, 1961.

\bibitem[{Howard}(1961)]{Howard61}
L.~N. {Howard}.
\newblock {Note on a paper of John W. Miles}.
\newblock \emph{JFM}, 10:\penalty0 509--512, 1961.

\bibitem[{Townsend}(1958)]{Townsend58}
A.~A. {Townsend}.
\newblock {The effects of radiative transfer on turbulent flow of a stratified
  fluid}.
\newblock \emph{JFM}, 4:\penalty0 361--375, 1958.

\bibitem[{Rayleigh}(1917)]{Rayleigh1917}
L.~{Rayleigh}.
\newblock {On the Dynamics of Revolving Fluids}.
\newblock \emph{Proceedings of the Royal Society of London Series A},
  93:\penalty0 148--154, March 1917.
\newblock \doi{10.1098/rspa.1917.0010}.

\bibitem[Solberg(1936)]{solberg1936mouvement}
H~Solberg.
\newblock Le mouvement d?inertie de l?atmosphere stable et son role dans la
  theorie des cyclones.
\newblock \emph{Proces-Verbaux des s{\'e}ances de l?Union International de
  G{\'e}od{\'e}sie et G{\'e}ophysique (IUGG)}, pages 66--82, 1936.

\bibitem[H{\o}iland(1941)]{hoiland1941stability}
Einar H{\o}iland.
\newblock \emph{On the Stability of the Circular Vortex.}
\newblock J. Dybwad, 1941.

\bibitem[{Goldreich} and {Schubert}(1967)]{GoldreichSchubert1967}
Peter {Goldreich} and Gerald {Schubert}.
\newblock {Differential Rotation in Stars}.
\newblock \emph{ApJ}, 150:\penalty0 571, November 1967.
\newblock \doi{10.1086/149360}.

\bibitem[{Spiegel} and {Zahn}(1970)]{SpiegelZahn1970}
E.~A. {Spiegel} and J.~P. {Zahn}.
\newblock {Instabilities of Differential Rotation}.
\newblock \emph{Comments on Astrophysics and Space Physics}, 2:\penalty0 178,
  September 1970.

\bibitem[{Dudis}(1974)]{Dudis1974}
J.~J. {Dudis}.
\newblock {The stability of a thermally radiating stratified shear layer,
  including self-absorption}.
\newblock \emph{JFM}, 64:\penalty0 65--83, 1974.

\bibitem[{Jones}(1977)]{Jones1977}
C.~A. {Jones}.
\newblock {The Onset of Shear Instability in Stars}.
\newblock \emph{Geophysical and Astrophysical Fluid Dynamics}, 8:\penalty0
  165--184, 1977.
\newblock \doi{10.1080/03091927708240377}.

\bibitem[{Prat} and {Ligni{\`e}res}(2013)]{PratLignieres13}
V.~{Prat} and F.~{Ligni{\`e}res}.
\newblock {Turbulent transport in radiative zones of stars}.
\newblock \emph{A.\&A.}, 551:\penalty0 L3, March 2013.

\bibitem[{Prat} and {Ligni{\`e}res}(2014)]{PratLignieres14}
V.~{Prat} and F.~{Ligni{\`e}res}.
\newblock {Shear mixing in stellar radiative zones. I. Effect of thermal
  diffusion and chemical stratification}.
\newblock \emph{{A.\&A.}}, 566:\penalty0 A110, June 2014.
\newblock \doi{10.1051/0004-6361/201423655}.

\bibitem[{Prat} et~al.(2016){Prat}, {Guilet}, {Viallet}, and
  {M{\"u}ller}]{Pratal2016}
V.~{Prat}, J.~{Guilet}, M.~{Viallet}, and E.~{M{\"u}ller}.
\newblock {Shear mixing in stellar radiative zones. II. Robustness of numerical
  simulations}.
\newblock \emph{{A.\&A.}}, 592:\penalty0 A59, July 2016.
\newblock \doi{10.1051/0004-6361/201527946}.

\bibitem[{Garaud} and {Kulenthirarajah}(2016)]{GaraudKulen16}
P.~{Garaud} and L.~{Kulenthirarajah}.
\newblock {Turbulent Transport in a Strongly Stratified Forced Shear Layer with
  Thermal Diffusion}.
\newblock \emph{ApJ}, 821:\penalty0 49, April 2016.
\newblock \doi{10.3847/0004-637X/821/1/49}.

\bibitem[{Garaud} et~al.(2017){Garaud}, {Gagnier}, and {Verhoeven}]{Garaudal17}
P.~{Garaud}, D.~{Gagnier}, and J.~{Verhoeven}.
\newblock {Turbulent Transport by Diffusive Stratified Shear Flows: From Local
  to Global Models. I. Numerical Simulations of a Stratified Plane Couette
  Flow}.
\newblock \emph{ApJ}, 837:\penalty0 133, March 2017.
\newblock \doi{10.3847/1538-4357/837/2/133}.

\bibitem[{Garaud} et~al.(2015{\natexlab{a}}){Garaud}, {Gallet}, and
  {Bischoff}]{Garaudal15}
P.~{Garaud}, B.~{Gallet}, and T.~{Bischoff}.
\newblock {The stability of stratified spatially periodic shear flows at low
  P{\'e}clet number}.
\newblock \emph{Physics of Fluids}, 27\penalty0 (8):\penalty0 084104, August
  2015{\natexlab{a}}.
\newblock \doi{10.1063/1.4928164}.

\bibitem[{Zahn}(1992)]{Zahn92}
J.-P. {Zahn}.
\newblock {Circulation and turbulence in rotating stars}.
\newblock \emph{A.\&A.}, 265:\penalty0 115--132, November 1992.

\bibitem[Billant and Chomaz(2001)]{BillantChomaz2001}
Paul Billant and Jean-Marc Chomaz.
\newblock Self-similarity of strongly stratified inviscid flows.
\newblock \emph{Physics of Fluids}, 13\penalty0 (6):\penalty0 1645--1651, 2001.
\newblock \doi{10.1063/1.1369125}.

\bibitem[{Brethouwer} et~al.(2007){Brethouwer}, {Billant}, {Lindborg}, and
  {Chomaz}]{Brethouweral2007}
G.~{Brethouwer}, P.~{Billant}, E.~{Lindborg}, and J.~M. {Chomaz}.
\newblock {Scaling analysis and simulation of strongly stratified turbulent
  flows}.
\newblock \emph{JFM}, 585:\penalty0 343, August 2007.
\newblock \doi{10.1017/S0022112007006854}.

\bibitem[{Drazin} and {Reid}(2004)]{DrazinReid}
P.~G. {Drazin} and W.~H. {Reid}.
\newblock \emph{{Hydrodynamic Stability}}.
\newblock {Cambridge University Press}, September 2004.

\bibitem[{Park} et~al.(1994){Park}, {Whitehead}, and
  {Gnanadeskian}]{Parkal1994}
Y.~G. {Park}, J.~A. {Whitehead}, and A.~{Gnanadeskian}.
\newblock {Turbulent mixing in stratified fluids: layer formation and
  energetics}.
\newblock \emph{JFM}, 279:\penalty0 279--311, January 1994.
\newblock \doi{10.1017/S0022112094003915}.

\bibitem[{Holford} and {Linden}(1999)]{HolfordLinden1999}
Joanne~M. {Holford} and P.~F. {Linden}.
\newblock {Turbulent mixing in a stratified fluid}.
\newblock \emph{Dynamics of Atmospheres and Oceans}, 30\penalty0 (2):\penalty0
  173--198, December 1999.
\newblock \doi{10.1016/S0377-0265(99)00025-1}.

\bibitem[{Oglethorpe} et~al.(2013){Oglethorpe}, {Caulfield}, and
  {Woods}]{Oglethorpeal2013}
R.~L.~F. {Oglethorpe}, C.~P. {Caulfield}, and Andrew~W. {Woods}.
\newblock {Spontaneous layering in stratified turbulent Taylor-Couette flow}.
\newblock \emph{JFM}, 721:\penalty0 R3, April 2013.
\newblock \doi{10.1017/jfm.2013.85}.

\bibitem[{Thorpe}(2016)]{Thorpe2016}
S.~A. {Thorpe}.
\newblock {Layers and internal waves in uniformly stratified fluids stirred by
  vertical grids}.
\newblock \emph{JFM}, 793:\penalty0 380--413, April 2016.
\newblock \doi{10.1017/jfm.2016.121}.

\bibitem[{Lucas} et~al.(2017){Lucas}, {Caulfield}, and {Kerswell}]{Lucasal2017}
Dan {Lucas}, C.~P. {Caulfield}, and Rich~R. {Kerswell}.
\newblock {Layer formation in horizontally forced stratified turbulence:
  connecting exact coherent structures to linear instabilities}.
\newblock \emph{JFM}, 832:\penalty0 409--437, December 2017.
\newblock \doi{10.1017/jfm.2017.661}.

\bibitem[Billant and Chomaz(2000)]{BillantChomaz2000}
P.~Billant and {J.-M.} Chomaz.
\newblock Experimental evidence for a new instability of a vertical columnar
  vortex pair in a strongly stratified fluid.
\newblock \emph{JFM}, 418:\penalty0 167--188, 2000.

\bibitem[{Phillips}(1972)]{Phillips1972}
O.~M. {Phillips}.
\newblock {Turbulence in a strongly stratified fluid -- is it unstable?}
\newblock \emph{Deep Sea Research and Oceanographic Abstracts}, 19:\penalty0
  79--81, 1972.
\newblock \doi{10.1016/0011-7471(72)90074-5}.

\bibitem[{Balmforth} et~al.(1998){Balmforth}, {Smith}, and {Young}]{ballsy}
N.~J. {Balmforth}, S.~G.~L. {Smith}, and W.~R. {Young}.
\newblock {Dynamics of interfaces and layers in a stratified turbulent fluid}.
\newblock \emph{JFM}, 355:\penalty0 329--358, January 1998.

\bibitem[{Caulfield}(2021)]{Caulfield2021}
C.~P. {Caulfield}.
\newblock {Layering, Instabilities, and Mixing in Turbulent Stratified Flows}.
\newblock \emph{Annual Review of Fluid Mechanics}, 53\penalty0 (1):\penalty0
  042320-100458, January 2021.
\newblock \doi{10.1146/annurev-fluid-042320-100458}.

\bibitem[{Ligni{\`e}res}(2020)]{Lignieres2020}
Fran{\c{c}}ois {Ligni{\`e}res}.
\newblock {Turbulence in stably stratified radiative zone}.
\newblock In Michel {Rieutord}, Isabelle {Baraffe}, and Yveline {Lebreton},
  editors, \emph{Multi-Dimensional Processes In Stellar Physics}, page 111,
  January 2020.

\bibitem[{Cope} et~al.(2020){Cope}, {Garaud}, and {Caulfield}]{Copeal20}
L.~{Cope}, P.~{Garaud}, and C.~P. {Caulfield}.
\newblock {The dynamics of stratified horizontal shear flows at low P{\'e}clet
  number}.
\newblock \emph{arXiv e-prints}, art. arXiv:1911.09674, November 2020.

\bibitem[{Garaud}(2020{\natexlab{b}})]{Garaud20}
P.~{Garaud}.
\newblock {Horizontal shear instabilities at low Prandtl number}.
\newblock \emph{arXiv e-prints}, art. arXiv:2006.07436, June
  2020{\natexlab{b}}.

\bibitem[{Stommel} et~al.(1956){Stommel}, {Arons}, and
  {Blanchard}]{Stommelal1956}
Henry {Stommel}, Arnold~B. {Arons}, and Duncan {Blanchard}.
\newblock {An oceanographical curiosity: the perpetual salt fountain}.
\newblock \emph{Deep Sea Research}, 3\penalty0 (2):\penalty0 152--153, February
  1956.
\newblock \doi{10.1016/0146-6313(56)90095-8}.

\bibitem[Stern(1960)]{stern1960sfa}
ME~Stern.
\newblock {The salt fountain and thermohaline convection}.
\newblock \emph{Tellus}, 12\penalty0 (2):\penalty0 172--175, 1960.

\bibitem[{Holyer}(1983)]{Holyer1983}
J.~Y. {Holyer}.
\newblock {Double-diffusive interleaving due to horizontal gradients}.
\newblock \emph{JFM}, 137:\penalty0 347--362, December 1983.
\newblock \doi{10.1017/S002211208300244X}.

\bibitem[{Radko}(2016)]{Radko2016}
T.~{Radko}.
\newblock {Thermohaline layering in dynamically and diffusively stable shear
  flows}.
\newblock \emph{JFM}, 805:\penalty0 147--170, October 2016.
\newblock \doi{10.1017/jfm.2016.547}.

\bibitem[Baines and Gill(1969)]{baines1969}
PG~Baines and AE~Gill.
\newblock {On thermohaline convection with linear gradients}.
\newblock \emph{JFM}, 37:\penalty0 289--306, 1969.

\bibitem[Radko(2013)]{radko2013double}
Timour Radko.
\newblock \emph{Double-diffusive convection}.
\newblock Cambridge University Press, 2013.

\bibitem[{Walin}(1964)]{Walin1964}
G.~{Walin}.
\newblock {Note on the stability of water stratified by both salt and heat}.
\newblock \emph{Tellus}, 16:\penalty0 389, August 1964.

\bibitem[Vauclair(2004)]{Vauclair04}
S.~Vauclair.
\newblock {Metallic Fingers and Metallicity Excess in Exoplanets' Host Stars:
  The Accretion Hypothesis Revisited}.
\newblock \emph{ApJ}, 605\penalty0 (2):\penalty0 874--879, 2004.

\bibitem[{Ulrich}(1972)]{Ulrich1972}
R.~K. {Ulrich}.
\newblock {Thermohaline Convection in Stellar Interiors.}
\newblock \emph{ApJ}, 172:\penalty0 165--+, February 1972.
\newblock \doi{10.1086/151336}.

\bibitem[{Kato}(1966)]{Kato1966}
S.~{Kato}.
\newblock {Overstable Convection in a Medium Stratified in Mean Molecular
  Weight}.
\newblock \emph{Proc. Astro. Soc. Japan}, 18:\penalty0 374, 1966.

\bibitem[{Spiegel}(1969)]{Spiegel1969}
E.~A. {Spiegel}.
\newblock {Semiconvection}.
\newblock \emph{Comments on Astrophysics and Space Physics}, 1:\penalty0 57,
  March 1969.

\bibitem[{Schwarzschild} and {H{\"a}rm}(1958)]{SchwarzschildHarm1958}
M.~{Schwarzschild} and R.~{H{\"a}rm}.
\newblock {Evolution of Very Massive Stars.}
\newblock \emph{ApJ}, 128:\penalty0 348, September 1958.
\newblock \doi{10.1086/146548}.

\bibitem[{Joseph}(1976)]{Joseph1976}
D.~D. {Joseph}.
\newblock {Stability of fluid motions. II}.
\newblock \emph{NASA STI/Recon Technical Report A}, 27:\penalty0 12423,
  November 1976.

\bibitem[{Balmforth} et~al.(2006){Balmforth}, {Ghadge}, {Kettapun}, and
  {Mandre}]{Balmforthal2006}
Neil~J. {Balmforth}, Shilpa~A. {Ghadge}, Atichart {Kettapun}, and Shreyas~D.
  {Mandre}.
\newblock {Bounds on double-diffusive convection}.
\newblock \emph{JFM}, 569:\penalty0 29--50, December 2006.
\newblock \doi{10.1017/S0022112006002230}.

\bibitem[{Schmitt}(1983)]{schmitt1983}
R.~W. {Schmitt}.
\newblock {The characteristics of salt fingers in a variety of fluid systems,
  including stellar interiors, liquid metals, oceans, and magmas}.
\newblock \emph{Phys. Fluids}, 26:\penalty0 2373--2377, September 1983.
\newblock \doi{10.1063/1.864419}.

\bibitem[{Radko} and {Smith}(2012)]{RadkoSmith2012}
T.~{Radko} and D.~P. {Smith}.
\newblock {Equilibrium transport in double-diffusive convection}.
\newblock \emph{JFM}, 692:\penalty0 5--27, February 2012.
\newblock \doi{10.1017/jfm.2011.343}.

\bibitem[{Brown} et~al.(2013){Brown}, {Garaud}, and {Stellmach}]{Brownal2013}
J.~M. {Brown}, P.~{Garaud}, and S.~{Stellmach}.
\newblock {Chemical Transport and Spontaneous Layer Formation in Fingering
  Convection in Astrophysics}.
\newblock \emph{ApJ}, 768:\penalty0 34, May 2013.
\newblock \doi{10.1088/0004-637X/768/1/34}.

\bibitem[{Denissenkov}(2010)]{denissenkov2010}
P.~A. {Denissenkov}.
\newblock {Numerical Simulations of Thermohaline Convection: Implications for
  Extra-mixing in Low-mass RGB Stars}.
\newblock \emph{ApJ}, 723:\penalty0 563--579, November 2010.
\newblock \doi{10.1088/0004-637X/723/1/563}.

\bibitem[Tait and Howe(1968)]{tait1968sot}
R.I. Tait and M.R. Howe.
\newblock {Some observations of thermohaline stratification in the deep ocean}.
\newblock \emph{Deep Sea Res.}, 15:\penalty0 275--280, 1968.

\bibitem[Tait and Howe(1971)]{tait1971ts}
R.I. Tait and M.R. Howe.
\newblock {Thermohaline staircase}.
\newblock \emph{Nature}, 231\penalty0 (5299):\penalty0 178--179, 1971.

\bibitem[Schmitt et~al.(1987)Schmitt, Perkins, Boyd, and Stalcup]{schmitt1987c}
Raymond~W Schmitt, H~Perkins, JD~Boyd, and MC~Stalcup.
\newblock C-salt: an investigation of the thermohaline staircase in the western
  tropical north atlantic.
\newblock \emph{Deep Sea Res.}, 34\penalty0 (10):\penalty0 1655--1665, 1987.

\bibitem[Krishnamurti(2003)]{krishnamurti2003double}
R~Krishnamurti.
\newblock Double-diffusive transport in laboratory thermohaline staircases.
\newblock \emph{JFM}, 483:\penalty0 287--314, 2003.

\bibitem[Krishnamurti(2009)]{krishnamurti2009heat}
R~Krishnamurti.
\newblock Heat, salt and momentum transport in a laboratory thermohaline
  staircase.
\newblock \emph{JFM}, 638:\penalty0 491--506, 2009.

\bibitem[Radko(2003)]{radko2003mechanism}
Timour Radko.
\newblock A mechanism for layer formation in a double-diffusive fluid.
\newblock \emph{JFM}, 497:\penalty0 365--380, 2003.

\bibitem[{Stellmach} et~al.(2011){Stellmach}, {Traxler}, {Garaud}, {Brummell},
  and {Radko}]{Stellmach2011}
S.~{Stellmach}, A.~{Traxler}, P.~{Garaud}, N.~{Brummell}, and T.~{Radko}.
\newblock {Dynamics of fingering convection. Part 2 The formation of
  thermohaline staircases}.
\newblock \emph{JFM}, 677:\penalty0 554--571, June 2011.
\newblock \doi{10.1017/jfm.2011.99}.

\bibitem[Schmitt et~al.(2005)Schmitt, Ledwell, Montgomery, Polzin, and
  Toole]{schmitt2005enhanced}
Raymond~W Schmitt, JR~Ledwell, ET~Montgomery, KL~Polzin, and JM~Toole.
\newblock Enhanced diapycnal mixing by salt fingers in the thermocline of the
  tropical atlantic.
\newblock \emph{Science}, 308\penalty0 (5722):\penalty0 685--688, 2005.

\bibitem[{Radko}(2010)]{Radko2010}
Timour {Radko}.
\newblock {Equilibration of weakly nonlinear salt fingers}.
\newblock \emph{JFM}, 645:\penalty0 121, February 2010.
\newblock \doi{10.1017/S0022112009992552}.

\bibitem[Xie et~al.(2017)Xie, Miquel, Julien, and Knobloch]{xie2017reduced}
Jin-Han Xie, Benjamin Miquel, Keith Julien, and Edgar Knobloch.
\newblock A reduced model for salt-finger convection in the small diffusivity
  ratio limit.
\newblock \emph{Fluids}, 2\penalty0 (1):\penalty0 6, 2017.

\bibitem[{Traxler} et~al.(2011){Traxler}, {Garaud}, and
  {Stellmach}]{Traxler2011b}
A.~{Traxler}, P.~{Garaud}, and S.~{Stellmach}.
\newblock {Numerically Determined Transport Laws for Fingering
  (''Thermohaline'') Convection in Astrophysics}.
\newblock \emph{ApJL}, 728:\penalty0 L29, February 2011.
\newblock \doi{10.1088/2041-8205/728/2/L29}.

\bibitem[{Garaud} et~al.(2015{\natexlab{b}}){Garaud}, {Medrano}, {Brown},
  {Mankovich}, and {Moore}]{Garaudal15b}
P.~{Garaud}, M.~{Medrano}, J.~M. {Brown}, C.~{Mankovich}, and K.~{Moore}.
\newblock {Excitation of Gravity Waves by Fingering Convection, and the
  Formation of Compositional Staircases in Stellar Interiors}.
\newblock \emph{ApJ}, 808:\penalty0 89, July 2015{\natexlab{b}}.
\newblock \doi{10.1088/0004-637X/808/1/89}.

\bibitem[{Kippenhahn} et~al.(1980){Kippenhahn}, {Ruschenplatt}, and
  {Thomas}]{kippenhahn80}
R.~{Kippenhahn}, G.~{Ruschenplatt}, and {H.-C.} {Thomas}.
\newblock {The time scale of thermohaline mixing in stars}.
\newblock \emph{A.\&A.}, 91:\penalty0 175--180, November 1980.

\bibitem[{Garaud} et~al.(2019){Garaud}, {Kumar}, and {Sridhar}]{Garaudal2019}
P.~{Garaud}, A.~{Kumar}, and J.~{Sridhar}.
\newblock {The Interaction between Shear and Fingering (Thermohaline)
  Convection}.
\newblock \emph{ApJ}, 879\penalty0 (1):\penalty0 60, July 2019.
\newblock \doi{10.3847/1538-4357/ab232f}.

\bibitem[Veronis et~al.(1965)]{veronis1965finite}
George Veronis et~al.
\newblock On finite amplitude instability in thermohaline convection.
\newblock \emph{J. Mar. Res}, 23\penalty0 (1):\penalty0 1--17, 1965.

\bibitem[{Proctor}(1981)]{Proctor1981}
M.~R.~E. {Proctor}.
\newblock {Steady subcritical thermohaline convection}.
\newblock \emph{JFM}, 105:\penalty0 507--521, April 1981.
\newblock \doi{10.1017/S0022112081003315}.

\bibitem[Turner(1965)]{TURNER1965}
J.S. Turner.
\newblock The coupled turbulent transports of salt and and heat across a sharp
  density interface.
\newblock \emph{International Journal of Heat and Mass Transfer}, 8\penalty0
  (5):\penalty0 759 -- 767, 1965.
\newblock ISSN 0017-9310.
\newblock \doi{http://dx.doi.org/10.1016/0017-9310(65)90022-0}.
\newblock URL
  \url{http://www.sciencedirect.com/science/article/pii/0017931065900220}.

\bibitem[{Linden} and {Shirtcliffe}(1978)]{LindenShirtcliffe1978}
P.~F. {Linden} and T.~G.~L. {Shirtcliffe}.
\newblock {The diffusive interface in double-diffusive convection}.
\newblock \emph{JFM}, 87:\penalty0 417--432, 1978.
\newblock \doi{10.1017/S002211207800169X}.

\bibitem[Carpenter et~al.(2012)Carpenter, Sommer, and
  W{\"u}est]{carpenter2012simulations}
JR~Carpenter, T~Sommer, and A~W{\"u}est.
\newblock Simulations of a double-diffusive interface in the diffusive
  convection regime.
\newblock \emph{JFM}, 711:\penalty0 411--436, 2012.

\bibitem[{Timmermans} et~al.(2008){Timmermans}, {Toole}, {Krishfield}, and
  {Winsor}]{Timmermans2008}
M.-L. {Timmermans}, J.~{Toole}, R.~{Krishfield}, and P.~{Winsor}.
\newblock {Ice-Tethered Profiler observations of the double-diffusive staircase
  in the Canada Basin thermocline}.
\newblock \emph{Journal of Geophysical Research (Oceans)}, 113:\penalty0
  C00A02, January 2008.
\newblock \doi{10.1029/2008JC004829}.

\bibitem[{W\"uest} et~al.(2012){W\"uest}, {Sommer}, and {Carpenter}]{Wuest2012}
A.~{W\"uest}, T.~{Sommer}, and J.~R. {Carpenter}.
\newblock \emph{{Diffusive-type of double diffusion in lakes: a review}}, pages
  271--284.
\newblock IAHR Monographs, CRC Press / Taylor \&Francis Group, 2012.

\bibitem[{Stevenson}(1977)]{Stevenson1977}
D.~J. {Stevenson}.
\newblock {A Semitheory for Semiconvection}.
\newblock \emph{Proceedings of the Astronomical Society of Australia}, 3,
  September 1977.

\bibitem[{Langer} et~al.(1983){Langer}, {Fricke}, and {Sugimoto}]{Langer1983}
N.~{Langer}, K.~J. {Fricke}, and D.~{Sugimoto}.
\newblock {Semiconvective diffusion and energy transport}.
\newblock \emph{A.\&A.}, 126:\penalty0 207, September 1983.

\bibitem[{Mirouh} et~al.(2012){Mirouh}, {Garaud}, {Stellmach}, {Traxler}, and
  {Wood}]{Mirouh2012}
G.~M. {Mirouh}, P.~{Garaud}, S.~{Stellmach}, A.~L. {Traxler}, and T.~S. {Wood}.
\newblock {A New Model for Mixing by Double-diffusive Convection
  (Semi-convection). I. The Conditions for Layer Formation}.
\newblock \emph{ApJ}, 750:\penalty0 61, May 2012.
\newblock \doi{10.1088/0004-637X/750/1/61}.

\bibitem[{Paparella} et~al.(2002){Paparella}, {Spiegel}, and
  {Talon}]{Paparella2002}
F.~{Paparella}, E.~A. {Spiegel}, and S.~{Talon}.
\newblock {Shear and Mixing in Oscillatory Doubly Diffusive Convection}.
\newblock \emph{Geophysical and Astrophysical Fluid Dynamics}, 96:\penalty0
  271--289, April 2002.
\newblock \doi{10.1080/03091920290029031}.

\bibitem[{Moll} et~al.(2016){Moll}, {Garaud}, and {Stellmach}]{Moll2016}
R.~{Moll}, P.~{Garaud}, and S.~{Stellmach}.
\newblock {A New Model for Mixing by Double-diffusive Convection
  (Semi-convection). III. Thermal and Compositional Transport through
  Non-layered ODDC}.
\newblock \emph{ApJ}, 823:\penalty0 33, May 2016.
\newblock \doi{10.3847/0004-637X/823/1/33}.

\bibitem[{Rosenblum} et~al.(2011){Rosenblum}, {Garaud}, {Traxler}, and
  {Stellmach}]{rosenblumal2011}
E.~{Rosenblum}, P.~{Garaud}, A.~{Traxler}, and S.~{Stellmach}.
\newblock {Turbulent Mixing and Layer Formation in Double-diffusive Convection:
  Three-dimensional Numerical Simulations and Theory}.
\newblock \emph{ApJ}, 731:\penalty0 66, April 2011.
\newblock \doi{10.1088/0004-637X/731/1/66}.

\bibitem[{Moore} and {Garaud}(2016)]{MooreGaraud2016}
K.~{Moore} and P.~{Garaud}.
\newblock {Main Sequence Evolution with Layered Semiconvection}.
\newblock \emph{ApJ}, 817:\penalty0 54, January 2016.
\newblock \doi{10.3847/0004-637X/817/1/54}.

\bibitem[{Wood} et~al.(2013){Wood}, {Garaud}, and {Stellmach}]{Woodal13}
T.~S. {Wood}, P.~{Garaud}, and S.~{Stellmach}.
\newblock {A New Model for Mixing by Double-diffusive Convection
  (Semi-convection). II. The Transport of Heat and Composition through Layers}.
\newblock \emph{ApJ}, 768:\penalty0 157, May 2013.
\newblock \doi{10.1088/0004-637X/768/2/157}.

\bibitem[{Moll} et~al.(2017){Moll}, {Garaud}, {Mankovich}, and
  {Fortney}]{Mollal2017}
R.~{Moll}, P.~{Garaud}, C.~{Mankovich}, and J.~J. {Fortney}.
\newblock {Double-diffusive Erosion of the Core of Jupiter}.
\newblock \emph{ApJ}, 849\penalty0 (1):\penalty0 24, November 2017.
\newblock \doi{10.3847/1538-4357/aa8d74}.

\bibitem[{Biello}(2001)]{Biello2001}
Joseph~Anthony {Biello}.
\newblock \emph{{Layer formation in semiconvection}}.
\newblock PhD thesis, THE UNIVERSITY OF CHICAGO, June 2001.

\bibitem[{Zaussinger} and {Spruit}(2013)]{ZaussingerSpruit2013}
F.~{Zaussinger} and H.~C. {Spruit}.
\newblock {Semiconvection: numerical simulations}.
\newblock \emph{A.\&A.}, 554:\penalty0 A119, June 2013.
\newblock \doi{10.1051/0004-6361/201220573}.

\bibitem[{Zaussinger} and {Kupka}(2019)]{ZaussingerKupka2019}
Florian {Zaussinger} and Friedrich {Kupka}.
\newblock {Layer formation in double-diffusive convection over resting and
  moving heated plates}.
\newblock \emph{Theoretical and Computational Fluid Dynamics}, 33\penalty0
  (3-4):\penalty0 383--409, August 2019.
\newblock \doi{10.1007/s00162-019-00499-7}.

\bibitem[{Stevenson} and {Salpeter}(1977)]{StevensonSalpeter1977}
D.~J. {Stevenson} and E.~E. {Salpeter}.
\newblock {The dynamics and helium distribution in hydrogen-helium fluid
  planets}.
\newblock \emph{ApJS}, 35:\penalty0 239--261, October 1977.
\newblock \doi{10.1086/190479}.

\bibitem[{Guillot} et~al.(2004){Guillot}, {Stevenson}, {Hubbard}, and
  {Saumon}]{Guillot2004}
T.~{Guillot}, D.~J. {Stevenson}, W.~B. {Hubbard}, and D.~{Saumon}.
\newblock \emph{{The interior of Jupiter}}, pages 35--57.
\newblock 2004.

\bibitem[{Cheng} et~al.(2018){Cheng}, {Aurnou}, {Julien}, and
  {Kunnen}]{Chengal2018}
Jonathan~S. {Cheng}, Jonathan~M. {Aurnou}, Keith {Julien}, and Rudie P.~J.
  {Kunnen}.
\newblock {A heuristic framework for next-generation models of geostrophic
  convective turbulence}.
\newblock \emph{Geophysical and Astrophysical Fluid Dynamics}, 112\penalty0
  (4):\penalty0 277--300, July 2018.
\newblock \doi{10.1080/03091929.2018.1506024}.

\bibitem[{Plumley} and {Julien}(2019)]{PlumleyJulien2019}
Meredith {Plumley} and Keith {Julien}.
\newblock {Scaling Laws in Rayleigh-B{\'e}nard Convection}.
\newblock \emph{Earth and Space Science}, 6\penalty0 (9):\penalty0 1580--1592,
  September 2019.
\newblock \doi{10.1029/2019EA000583}.

\bibitem[{Sengupta} and {Garaud}(2018)]{SenguptaGaraud2018}
S.~{Sengupta} and P.~{Garaud}.
\newblock {The Effect of Rotation on Fingering Convection in Stellar
  Interiors}.
\newblock \emph{ApJ}, 862\penalty0 (2):\penalty0 136, August 2018.
\newblock \doi{10.3847/1538-4357/aacbc8}.

\bibitem[{Moll} and {Garaud}(2017)]{MollGaraud2017}
R.~{Moll} and P.~{Garaud}.
\newblock {The Effect of Rotation on Oscillatory Double-diffusive Convection
  (Semiconvection)}.
\newblock \emph{ApJ}, 834:\penalty0 44, January 2017.
\newblock \doi{10.3847/1538-4357/834/1/44}.

\bibitem[{Fricke}(1968)]{Fricke1968}
K.~{Fricke}.
\newblock {Instabilit{\"a}t station{\"a}rer Rotation in Sternen}.
\newblock \emph{Zeit. Astrophys.}, 68:\penalty0 317, January 1968.

\bibitem[{Barker} et~al.(2019){Barker}, {Jones}, and {Tobias}]{Barkeral2019}
A.~J. {Barker}, C.~A. {Jones}, and S.~M. {Tobias}.
\newblock {Angular momentum transport by the GSF instability: non-linear
  simulations at the equator}.
\newblock \emph{MNRAS}, 487\penalty0 (2):\penalty0 1777--1794, August 2019.
\newblock \doi{10.1093/mnras/stz1386}.

\bibitem[{Barker} et~al.(){Barker}, {Jones}, and {Tobias}]{Barkeral2020}
A.~J. {Barker}, C.~A. {Jones}, and S.~M. {Tobias}.
\newblock {Angular momentum transport, layering, and zonal jet formation by the
  GSF instability: non-linear simulations at a general latitude}.
\newblock \emph{MNRAS}.

\bibitem[{Rashid} et~al.(2008){Rashid}, {Jones}, and {Tobias}]{Rashidal2008}
F.~Q. {Rashid}, C.~A. {Jones}, and S.~M. {Tobias}.
\newblock {Hydrodynamic instabilities in the solar tachocline}.
\newblock \emph{A\&A}, 488\penalty0 (3):\penalty0 819--827, September 2008.
\newblock \doi{10.1051/0004-6361:200810039}.

\bibitem[{Park} et~al.(2020){Park}, {Prat}, and {Mathis}]{Parkal2020}
J.~{Park}, V.~{Prat}, and S.~{Mathis}.
\newblock {Horizontal shear instabilities in rotating stellar radiation zones.
  I. Inflectional and inertial instabilities and the effects of thermal
  diffusion}.
\newblock \emph{{A.\&A.}}, 635:\penalty0 A133, March 2020.
\newblock \doi{10.1051/0004-6361/201936863}.

\bibitem[{Moffatt}(1978)]{Moffatt1978}
H.~K. {Moffatt}.
\newblock \emph{{Magnetic field generation in electrically conducting fluids}}.
\newblock 1978.

\bibitem[{Rincon}(2019)]{Rincon2019}
Fran{\c{c}}ois {Rincon}.
\newblock {Dynamo theories}.
\newblock \emph{Journal of Plasma Physics}, 85\penalty0 (4):\penalty0
  205850401, August 2019.
\newblock \doi{10.1017/S0022377819000539}.

\bibitem[{Yan} et~al.(2021){Yan}, {Tobias}, and {Calkins}]{Yanal2021}
M.~{Yan}, S.M. {Tobias}, and M.S. {Calkins}.
\newblock {Scaling behaviour of small-scale dynamos driven by
  Rayleigh-B{\'e}nard convection}.
\newblock \emph{JFM, {\it in press}}, 2021.

\bibitem[{Charbonnel} and {Zahn}(2007)]{CharbonnelZahn2007b}
C.~{Charbonnel} and J.~P. {Zahn}.
\newblock {Inhibition of thermohaline mixing by a magnetic field in Ap star
  descendants: implications for the Galactic evolution of $^{3}$He}.
\newblock \emph{A\&A}, 476\penalty0 (3):\penalty0 L29--L32, December 2007.
\newblock \doi{10.1051/0004-6361:20078740}.

\bibitem[{Harrington} and {Garaud}(2019)]{HarringtonGaraud2019}
Peter~Z. {Harrington} and Pascale {Garaud}.
\newblock {Enhanced Mixing in Magnetized Fingering Convection, and Implications
  for Red Giant Branch Stars}.
\newblock \emph{ApJ Lett.}, 870\penalty0 (1):\penalty0 L5, January 2019.
\newblock \doi{10.3847/2041-8213/aaf812}.

\bibitem[{Skoutnev} et~al.(2021){Skoutnev}, {Squire}, and
  {Bhattacharjee}]{Skoutneval2021}
V.~{Skoutnev}, J.~{Squire}, and A.~{Bhattacharjee}.
\newblock {Small-scale Dynamo in Stably Stratified Turbulence}.
\newblock \emph{ApJ}, 906\penalty0 (1):\penalty0 61, January 2021.
\newblock \doi{10.3847/1538-4357/abc8ee}.

\bibitem[{Tobias} et~al.(2007){Tobias}, {Diamond}, and {Hughes}]{TDH2007}
Steven~M. {Tobias}, Patrick~H. {Diamond}, and David~W. {Hughes}.
\newblock {{\ensuremath{\beta}}-Plane Magnetohydrodynamic Turbulence in the
  Solar Tachocline}.
\newblock \emph{ApJ Lett.}, 667\penalty0 (1):\penalty0 L113--L116, September
  2007.
\newblock \doi{10.1086/521978}.

\bibitem[{Gilman}(1983)]{Gilman1983}
P.~A. {Gilman}.
\newblock {Dynamically consistent nonlinear dynamos driven by convection in a
  rotating spherical shell. II - Dynamos with cycles and strong feedbacks}.
\newblock \emph{ApJ Supp.}, 53:\penalty0 243--268, October 1983.
\newblock \doi{10.1086/190891}.

\bibitem[{Varela} et~al.(2016){Varela}, {Strugarek}, and {Brun}]{Varela2016}
J.~{Varela}, A.~{Strugarek}, and A.~S. {Brun}.
\newblock {Characterizing the feedback of magnetic field on the differential
  rotation of solar-like stars}.
\newblock \emph{Advances in Space Research}, 58\penalty0 (8):\penalty0
  1507--1521, October 2016.
\newblock \doi{10.1016/j.asr.2016.06.032}.

\end{thebibliography}

\end{document}